\documentclass[12pt,superscriptaddress,showpacs,setstack,amssymb,amsfonts,nofootinbib,showpacs]{iopart}
\usepackage{graphicx}
\usepackage[breaklinks=true,colorlinks=true]{hyperref}
\usepackage{breakurl}
\usepackage{setspace}

\newcommand{\eV}{\, {\rm eV}}

\newcommand{\K}{\, {\rm K}}

\newcommand{\lr}[1]{ \left( #1 \right) }
\newcommand{\lrs}[1]{ \left[ #1 \right] }

\newcommand{\vev}[1]{ \langle \, #1 \, \rangle }

\newcommand{\re}{ {\rm Re} \, }

\renewcommand{\det}[1]{ {\rm det} \left( #1 \right) }

\newcommand{\eq}[1]{\begin{eqnarray}#1\end{eqnarray}}  

\newcommand{\beq}{\begin{eqnarray}}
\newcommand{\eeq}{\end{eqnarray}}

\newcommand{\calh}{{\cal H}}

\newcommand{\calz}{{\cal Z}}

\begin{document}

\title[Quantum phase transitions on the hexagonal lattice]{Quantum phase transitions on the hexagonal lattice}
\author{Dominik Smith$^1$, Pavel Buividovich$^2$,
Michael K\"orner$^1$, Maksim Ulybyshev$^3$ and Lorenz von Smekal$^1$}
\address{$^1$ Institut f\"ur Theoretische Physik, Justus-Liebig-Universit\"at, 35392 Giessen, Germany}
\address{$^2$ Theoretical Physics Division, Department of Mathematical Sciences, University of Liverpool, Liverpool L69 3BX, UK}
\address{$^3$ Institut f\"ur Theoretische Physik, Julius-Maximilians-Universit\"at, 97074 W\"urzburg, Germany}
\ead{dominiksmith81@gmail.com}

\begin{abstract}
Hubbard-type models on the hexagonal lattice are of great interest, as they provide realistic descriptions
of graphene and other related materials. Hybrid Monte Carlo simulations offer a first-principles approach to
study their phase structure. Here, we review the present status of our work in this direction.
\end{abstract}

\maketitle

\section{Introduction}
The Hubbard model, which describes fermionic
quasi particles with contact interactions, continues to be
of profound interest, as it remains the quintessential example of an interacting fermion system,
and can qualitatively describe many non-perturbative phenomena, such as dynamical mass-gap generation
or superconductivity. On the honeycomb lattice, extended versions
with varying on-site, nearest- and next-to-nearest-neighbor interactions have been predicted to
host a large variety of gapped phases, such as spin-density wave (SDW) and charge-density wave (CDW) phases, topological insulators and spontaneous
Kekul\'{e} distortions, many of which can be realized using ultra-cold atoms trapped in optical 
lattices \cite{Bermudez_2010,Mazza_2012} or other experimental techniques. Long-range interaction potentials realistically describe the physics of both mono- and bilayer graphene
\cite{PhysRevLett.106.236805,PhysRevX.8.031087}.
Determinantal Quantum Monte Carlo simulations following Blankenbecler, Scalapino
and Sugar (BSS) remain the method of choice for obtaining first-principles results for fermionic systems with contact interactions.
In recent years however, the Hybrid-Monte-Carlo (HMC) method, which has a long history in high-energy physics, has found increasing application
for the study of Hubbard-type models with off-site interactions, as it can efficiently simulate non-diagonal interaction matrices.

In this article, we review a number of different results which we obtained by applying HMC on the honeycomb lattice.
After presenting the basics of the numerical setup (section \ref{sec:setup}), we start with a discussion of the
extended Hubbard model with on-site ($U$) and nearest-neighbor ($V_1$) interactions (section \ref{sec:hubbphase}).
In this model, we studied the competition between SDW and CDW order and determined the location and critical properties of a phase
boundary separating a semimetal from a SDW phase \cite{Buividovich:2018hubb}. Next we turn to the realistic long-range
potential of graphene (section \ref{sec:graphenephase}), which exhibits a semimetal-SDW transition
in the strong-coupling regime. By studying its critical properties we found an inconsistency with the Gross-Neveu
universality class, which applies in the case of short-range interactions, and concluded that a conformal phase transition
provides a more natural explanation for our results \cite{Buividovich:2018crq}.
Third we discuss graphene at finite spin density (section \ref{sec:vhs}). We focus here on the topological neck-disrupting Lifshitz
transition, which occurs when the Fermi level traverses saddle points in the single-particle bands, and which is accompanied
by a logarithmic divergence of the density of states and correspondingly the ferromagnetic
susceptibility. We found that this divergence follows a powerlaw in the presence of interactions
and is driven by the disconnected rather than the connected susceptibility, indicating an instability towards
formation of ordered electronic phases \cite{Korner:2017qhf}. And finally we discuss graphene with hydrogen adatoms,
which can be modeled as vacant lattice sites (section \ref{sec:vacancies}). Here we showed that the pairwise Ruderman-Kittel-Kasuya-Yosida (RKKY) potential
between adatoms is strongly affected by inter-electron interactions, such that dimer formation is suppressed, and that certain
configurations of adatom superlattices are dynamically stabilized \cite{PhysRevB.96.165411}.
We end with a short outlook (section \ref{sec:outlook}).

\section{Numerical setup}
\label{sec:setup}
The starting point is the interacting tight-binding theory on the honeycomb lattice. In the most general
form which we consider, the Hamiltonian is given by:
\beq
\label{eq:tightbinding}
\hspace{-10mm}
\hat{\mathcal{H}} = - \kappa \sum_{\left \langle x, y \right \rangle, \sigma} \left( \hat{c}^{\dagger}_{x,\sigma} \hat{c}_{y,\sigma}
+ \textrm{h.c.}\right) + \frac{1}{2}\sum_{x,y}  \hat{q}_x V_{xy}\hat{q}_y + \sum_{x} m_s \left( \hat{c}^{\dagger}_{x,\uparrow} \hat{c}_{x,\uparrow}
- \hat{c}^{\dagger}_{x,\downarrow} \hat{c}_{x,\downarrow} \right) .
\eeq
Here $\kappa$ is the hopping energy, $\langle x, y\rangle$ denotes nearest-neighbor sites, $\sigma = \uparrow, \downarrow$ labels spin
directions, $\hat{q}_x =\hat{c}^{\dagger}_{x,\uparrow} \hat{c}_{x,\uparrow}+\hat{c}^{\dagger}_{x,\downarrow} \hat{c}_{x,\downarrow}-1$
is the electric charge operator and $m_s$ denotes a ``staggered'' mass, which has an opposite sign on the two triangular sublattices
and is added to remove zero modes (we will specify below when we used a non-zero $m_s$).
The creation- and annihilation operators satisfy the anticommutation relations
$\{ \hat{c}_{x,\sigma}, \hat{c}^{\dagger}_{y,\sigma'} \}= \delta_{x,y} \delta_{\sigma, \sigma'}$.
The matrix elements $V_{xy}$ can be chosen freely to describe different two-body potentials and are only
restricted by the condition that $V$ be positive definite (different choices are used in each of the projects discussed below).
The theoretical groundwork for HMC simulations of (\ref{eq:tightbinding}) was originally worked out in \cite{Brower:2012zd}.
We will present a compact summary here, which also takes some more recent developments into account.

HMC is based on the path-integral formulation of the (grand-canonical) partition function $\calz$.
To derive it, we start with a symmetric Suzuki-Trotter decomposition, which yields
\begin{eqnarray}
\calz = \Tr \left( e^{-\beta \hat{\mathcal{H}} } \right) =
\Tr \left( e^{-\delta_{\tau} \hat{\mathcal{H}}_{0}} e^{-\delta_{\tau} \hat{\mathcal{H}}_\mathrm{int} } e^{-\delta_{\tau} \hat{\mathcal{H}}_{0}} \dots \right) + O(\delta^2_{\tau})~.
\label{eq:Trotter}
\end{eqnarray}
Here we have separated the interacting and non-interacting contributions to (\ref{eq:tightbinding}) and introduced a finite
stepsize $\delta_\tau=\beta/N_t$ in Euclidean time.
We now apply two variable transformations. The first is a canonical transformation where creation- and annihilation operators
for spin-down particles are replaced by hole operators, and the sign of these is flipped on one sublattice,
i.e. a transformation of the form
\begin{eqnarray}
\hat{c}_{x, \uparrow}, \hat{c}^{\dagger}_{x, \uparrow} \to  \hat{a}_x, \hat{a}^{\dagger}_x,\quad\quad\quad
\hat{c}_{x, \downarrow}, \hat{c}^{\dagger}_{x, \downarrow} \to  \pm \hat{b}^{\dagger}_x, \pm \hat{b}_x~.
\label{eq:trafo1}
\end{eqnarray}
This step is necessary to avoid a fermion sign problem (i.e. an indefinite measure of the path integral), and can only
be applied on bipartite lattices. The second is a Fierz transformation, which is applied to the on-site interaction term and
which mixes charge and spin sectors. It has the form
\begin{eqnarray}
\label{eq:complexint}
\frac{V_{xx}}{2} \hat{q}_x^2 = \eta \frac{V_{xx}}{2} \hat{q}_x^2 - (1-\eta) \frac{V_{xx}}{2}(\hat{q}_x')^2 + V_{xx}(1-\eta)\,\hat{q}_x',
\end{eqnarray}
where $\hat{q}_x' = \hat{a}^{\dagger}_x \hat{a}_x + \hat{b}^{\dagger}_x \hat{b}_x$ is the spin-density operator and the constant $\eta$
can be chosen in the range $(0,1)$. As we will see below, this leads
to the complexification of the auxiliary fields which we introduce, and serves to maintain ergodicity of HMC trajectories in absence
of a staggered mass $m_s$.

Expressing $\calz$ as an integral over classical field variables, which can then be sampled
stochastically, requires removing the fermionic operators from
(\ref{eq:Trotter}). This can be done by a form of Gaussian integration if the exponentials contain only
bilinears. The fourth-power terms which appear in $\hat{\mathcal{H}}_\mathrm{int}$ must thus be removed,
which is done by a Hubbard-Stratonovich transformation.
We apply two different variants to interaction terms coupling to $\hat{q}_x$ and  $\hat{q}_x'$.
The first term $\sim \eta \hat{q}_x^2 $ appearing on the right hand side of (\ref{eq:complexint}) is re-absored
into the interaction matrix $V_{xy}$ and the combined expression is then transformed using
\begin{eqnarray}
\label{continuous_HS_imag}
  \hspace{-5mm}   \exp\left(-\frac{\delta_\tau}{2}
  \sum_{x,y}\hat q_x
   V_{xy} \hat q_y \right)   \cong \int D \phi \,
  \exp\left( - \frac{1}{2\delta_\tau} \sum_{x,y} \phi_x V^{-1}_{xy} \phi_y + i \delta_\tau \sum_x \phi_x\, \hat q_x \right).
\end{eqnarray}
The  $\sim (1-\eta)(\hat{q}_x')^2$ term is transformed by its own, using
\begin{eqnarray}
 \label{continuous_HS_real}
 \hspace{-15mm}  \exp\left(\frac{\delta_\tau}{2}(1-\eta) \sum_x V_{xx} (\hat q_x')^2 \right) \cong \int D \chi\, \exp\left(- \frac{1}{2\delta_\tau}
\sum_x \frac{\chi^2_x}{(1-\eta)V_{xx}} + \delta_\tau \sum_x \chi_x\, \hat q_x' \right).
\end{eqnarray}
In effect, we have introduced a complex auxiliary field variable (with real part $\chi$ and imaginary part $i\phi$)
at each site of the $2+1$ dimensional hypercubic lattice. The third term in (\ref{eq:complexint}) is absorbed into $\chi$
by the transformation $\chi \to\chi-\delta_\tau V_{xx}(1-\eta)$.

To compute the trace in the fermionic Fock space (with anti-periodic boundary conditions) appearing in (\ref{eq:Trotter}) we use
\begin{eqnarray}
\label{fermionic_identity}
 \Tr\left( e^{-\hat{A}_1 } e^{-\hat{A}_2 } \ldots e^{-\hat{A}_n } \right) = \det{
  \begin{array}{cccc}
     1          & -e^{-A_1} & 0        & \ldots  \\
     0          & 1        & -e^{-A_2} & \ldots  \\
        \vdots  &          & \ddots          &         \\
      e^{-A_n} & 0        & \ldots   & 1       \\
  \end{array}
 } ,
\end{eqnarray}
where $\hat{A}_k = \lr{A_k}_{ij} \hat{c}^{\dag}_i \hat{c}_j$ are the fermionic bilinear operators and $A_k$ (without hat) contain matrix elements in the single-particle Hilbert space (this identity is derived in ~\cite{Hirsch:85:1,Blankenbecler:81:1}).
We finally obtain
\begin{equation}
\label{eq:func_int}
  \calz = \int D \Phi \,|\textrm{det}[ M( \Phi)]|^2
  e^{-S_\eta(\Phi)},\,
\label{eq:part_func_int}
  \end{equation}
  with
  \begin{equation}
  S_\eta(\Phi) = \frac{1}{ 2  \delta_\tau}\sum_{x,y,t} \phi_{x,t}\widetilde{V}_{xy}^{-1}\phi_{y,t}
  + \sum_{x,t} \frac {(\chi_{x,t}- (1-\eta) \delta_\tau V_{xx})^2} {2 (1-\eta) \delta_\tau V_{xx}}.
  \label{eq:action_alpha}
\end{equation}
Here $\widetilde{V}$ denotes a modified interaction matrix wherein the diagonal elements have been 
rescaled by a factor of $\eta$ by (\ref{eq:complexint}). $|\textrm{det}[ M( \Phi)]|^2$
appears in (\ref{eq:part_func_int}) since, after the particle-hole transformation (\ref{eq:trafo1}), 
the spin-up and spin-down electrons contribute as $M$ and $M^\dagger$ respectively.

Two versions of the fermion matrix $M(\Phi)$ are used in this work which agree up to discretization effects.
The first a full exponential form which follows directly from (\ref{fermionic_identity}) and which respects the full
spin- and sublattice symmetries of the continuum limit but is non-sparse and must be inverted by a special Schur complement solver
(see \cite{Buividovich:2018hubb} for a discussion). The second is obtained by a linearization of terms $e^{-\delta_\tau h}$
where $h$ denotes the single-particle tight-binding hopping matrix. This form respects the above mentioned symmetries only in the time-continuum limit, but is sparse and thus can be inverted by a standard conjugate gradient solver
(see e.g. \cite{Smith:2014tha} and \cite{Korner:2017qhf}). As the expressions for the fermion matrices are rather
lengthy and differ between the works summarized in this review, we refer the interested reader to the
original publications. For a discussion of the relative advantages of different fermion discretizations,
see \cite{Buividovich:2016tgo} and \cite{PhysRevB.93.155106}.

The HMC algorithm essentially consists of combining the classical time evolution of $\Phi$ in computer time
with a stochastic refreshment of an associated momentum field $\pi$.
A stochastic representation of the fermion determinants
\begin{eqnarray}
\label{eq:stoch_det}
|\det M( \Phi)|^2 = \int D \Psi \Psi^\dag e^{-\Psi^\dag (M M^\dag)^{-1}\Psi },
 \end{eqnarray}
is used, which introduces an additional pseudofermionic field $\Psi$
(this is a regular complex-valued field, not a set of Grassmann variables). Ultimately, we must
numerically solve the Hamiltonian system given by
\begin{equation}
\hspace{-15mm}\label{eq:hmcmd}
\calh = S_\eta(\Phi)+\Psi^\dagger( M M^\dagger )^{-1} \Psi+ \frac{\pi^T \pi}{2}~,\quad
\left[ \frac{d \Phi}{d \tau} \right]^T= \frac{\partial \calh}{\partial \pi}~,
\quad
\left[ \frac{d \pi}{d \tau} \right]^T= - \frac{\partial \calh}{\partial \Phi}~.
\end{equation}
A symplectic integrator is employed, which introduces a stepsize error, which is then corrected
by a Metropolis accept/reject step. The full algorithm reads as follows:
\begin{samepage}
\begin{itemize}
\item[1)]{Refresh momentum field $\pi$ using Gaussian noise: $P(\pi)\sim e^{-\pi^2/2}$}~.
\item[2)]{Refresh pseudofermions $\Psi$ by generating field $\rho$ with
$P(\rho)=e^{-\rho^\dagger \rho}$ and obtaining $\Psi=M \rho$~.}
\item[3)]{Do Hamiltonian evolution of $\pi$ and $\Phi$ by solving (\ref{eq:hmcmd}) with symplectic
integrator (``molecular dynamics trajectory'').}
\item[4)]{Accept new configuration with probability $P=\min(1,e^{-\Delta \calh})$ (``Metropolis check'').}
\item[5)]{Continue from step 1).}
\end{itemize}
\end{samepage}
Additional details can be found in \cite{Smith:2014tha} and \cite{Buividovich:2018hubb}.

We point out here that a variant of HMC which avoids the
use of pseudofermions was described and used in \cite{Buividovich:2018crq}.
We also point out that we have omitted a discussion of boundary conditions. We used periodic boundary conditions
in space in each project discussed here, either of the Born-von K\'{a}rm\'{a}n type or with a rectangular geometry.
The exact type can be found in the original literature.

\section{Competing order in the extended Hubbard model}
\label{sec:hubbphase}
The hexagonal Hubbard model with pure on-site interaction has been studied fairly extensively with Quantum Monte
Carlo methods \cite{PhysRevX.3.031010,PhysRevX.6.011029,PhysRevB.91.165108}. By now it is well established
that it exhibits a second order phase transition from a semimetal to a gapped SDW phase if the onsite
potential is increased above $U\approx 3.8 \kappa$. The universality class of the $2+1D$ chiral Gross-Neveu model
has been verified for this transition and several critical exponents have been obtained with fairly high precision.
However, energy balance arguments and renormalization group calculations suggest that a CDW phase might be
favored in some region of the phase diagram if a nearest-neighbor potential $V_1$ is included (see e.g. \cite{PhysRevB.93.125119}).
To detect a particular phase one may use external sources.
For instance, the staggered mass $m_s$ introduced in (\ref{eq:tightbinding}) couples to an
anti-ferromagnetic condensate, and by extrapolating it to the $m_s\to 0$ and thermodynamic limits
one may probe the semimetal-SDW transition. To study competing phases one must add multiple
sources, so extrapolations in a high-dimensional parameter space are required. More importantly however,
a corresponding source term for CDW order introduces a sign problem and cannot be simulated, even in principle.

In \cite{Buividovich:2018hubb} we carried out an unbiased study of the competition between CDW and SDW
order in the $U-V_1$ plane by avoiding the use source terms altogether. We used
observables of the form
\begin{eqnarray}
\label{eq:squarespin}
 \langle X^2 \rangle
 =
 \left\langle \frac{1}{L^4} \left( \sum\limits_{x\in A} \hat{X}_{x} \right)^2 \right\rangle
 +
 \left\langle \frac{1}{L^4} \left( \sum\limits_{x\in B} \hat{X}_{x} \right)^2 \right\rangle ,
\end{eqnarray}
where the sums run over the two triangular sublattices (``A'' and ``B'') respectively and
$L$ is the linear lattice size. Observables of this type develop a non-zero expectation
value even without explicit symmetry breaking. We considered two different observables,
where $X$ is either the charge $q$ or a spin component $S_i$, with
\begin{eqnarray}
 \hat{S}_{x,i} = \frac{1}{2} ( \hat
 c^\dag_{x, \uparrow} , \, \hat c^\dag_{x, \downarrow} )  \sigma_i
 \left(
 \begin{array}{c}
  \hat{c}_{x, \uparrow} \\
  \hat{c}_{x, \downarrow} \\
 \end{array}
 \right).
\end{eqnarray}
Different spin components can be averaged over to improve statistics. We used the exponential
fermion operator in this project, so the rotational symmetry of $\hat{S}$ is respected.

We point out here that the Hamiltonian possesses zero modes in absence of any mass terms, which
partition the configuration space of a purely real or imaginary Hubbard field into
disconnected sectors, separated by infinite potential barriers. It is for this principle reason that
complex auxiliary fields were introduced in section \ref{sec:setup}. By choosing a mixing
parameter $\eta$ which is only slightly smaller than one, HMC trajectories are able to circumvent
these barriers, and ergodicity is recovered.

\begin{figure}
\begin{center}
\includegraphics[width=0.47\linewidth]{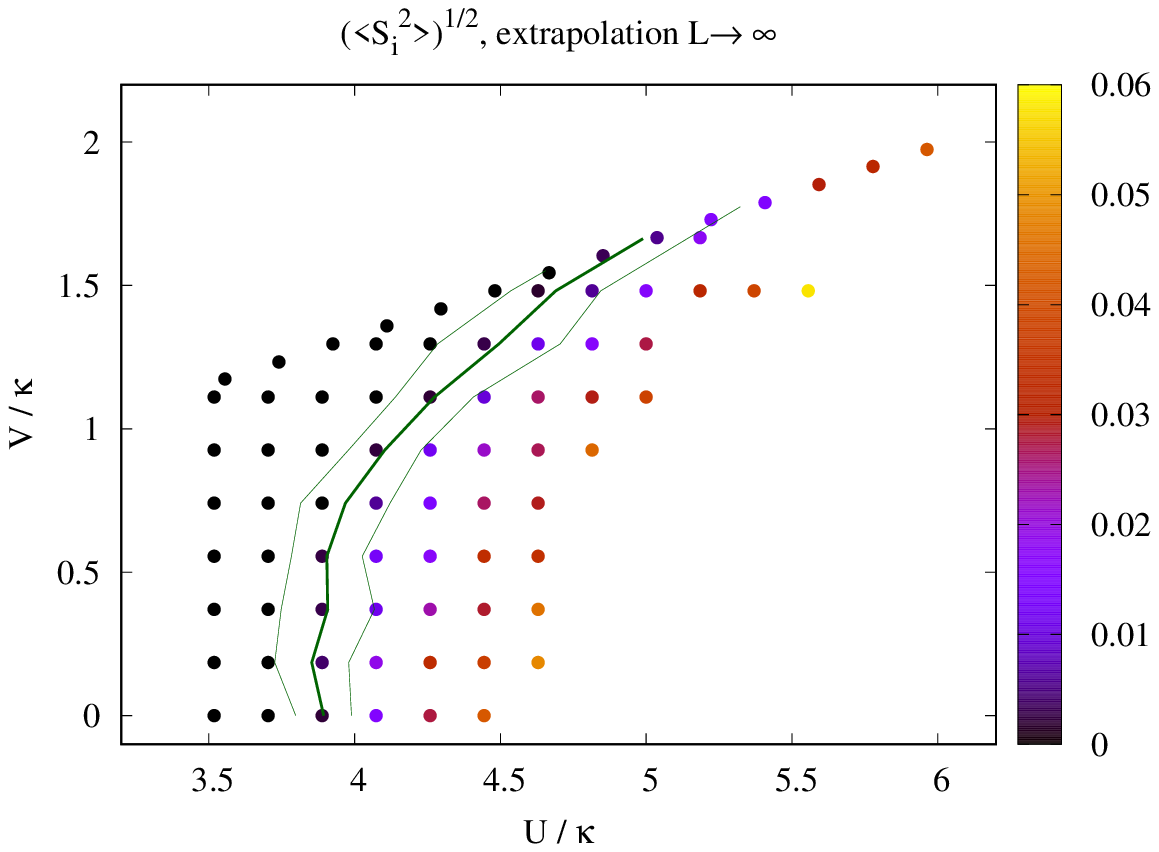}
\includegraphics[width=0.47\linewidth]{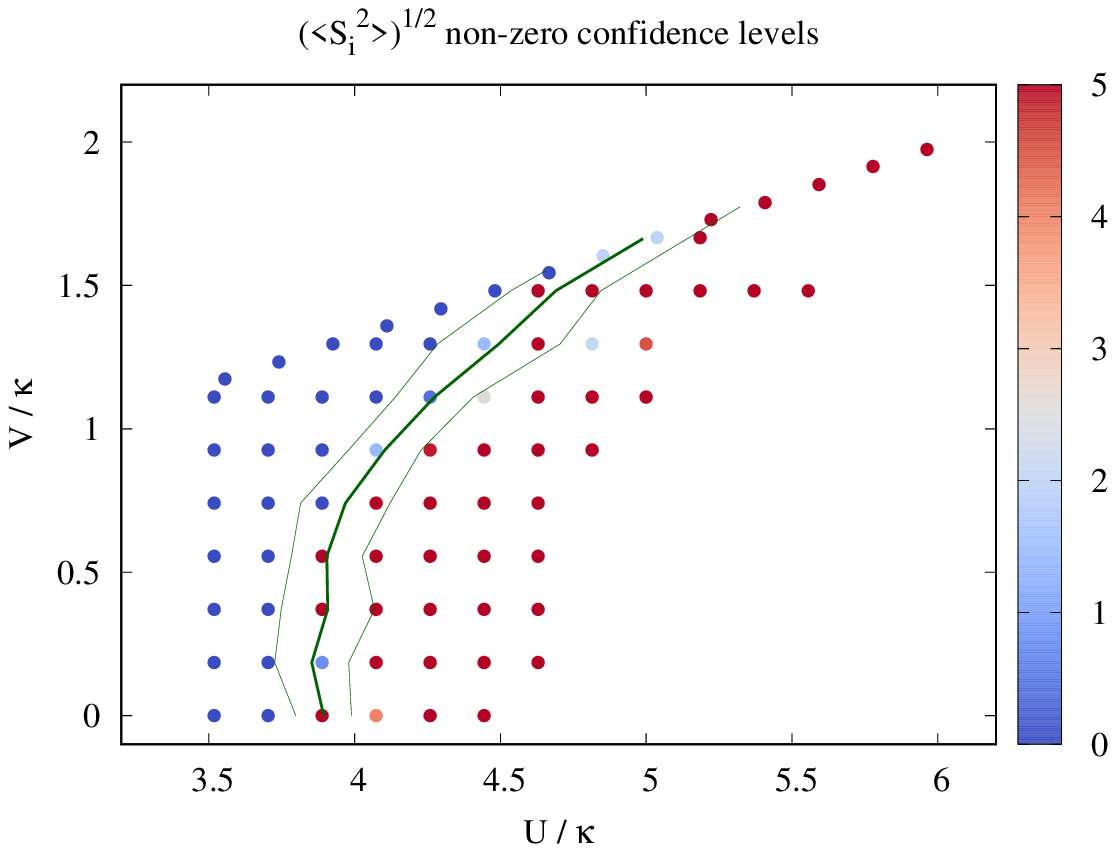}
\caption{ Figures taken from \cite{Buividovich:2018hubb}. \emph{Left:} Infinite volume extrapolation of $\sqrt{\langle S^2 \rangle}$ using linear model
$a+b\cdot(1/L)$. \emph{Right:} Number of standard deviations with which non-zero $\sqrt{\langle S^2 \rangle}$ is
established in $L\to \infty$ limit. \emph{Both figures:} Green lines mark the phase boundary with errorband, obtained from a
finite size scaling study of $\langle S^2 \rangle$.}
\label{fig:hubb_phase}
\end{center}
\end{figure}

\begin{figure}
\centering
\includegraphics[width=0.47\linewidth]{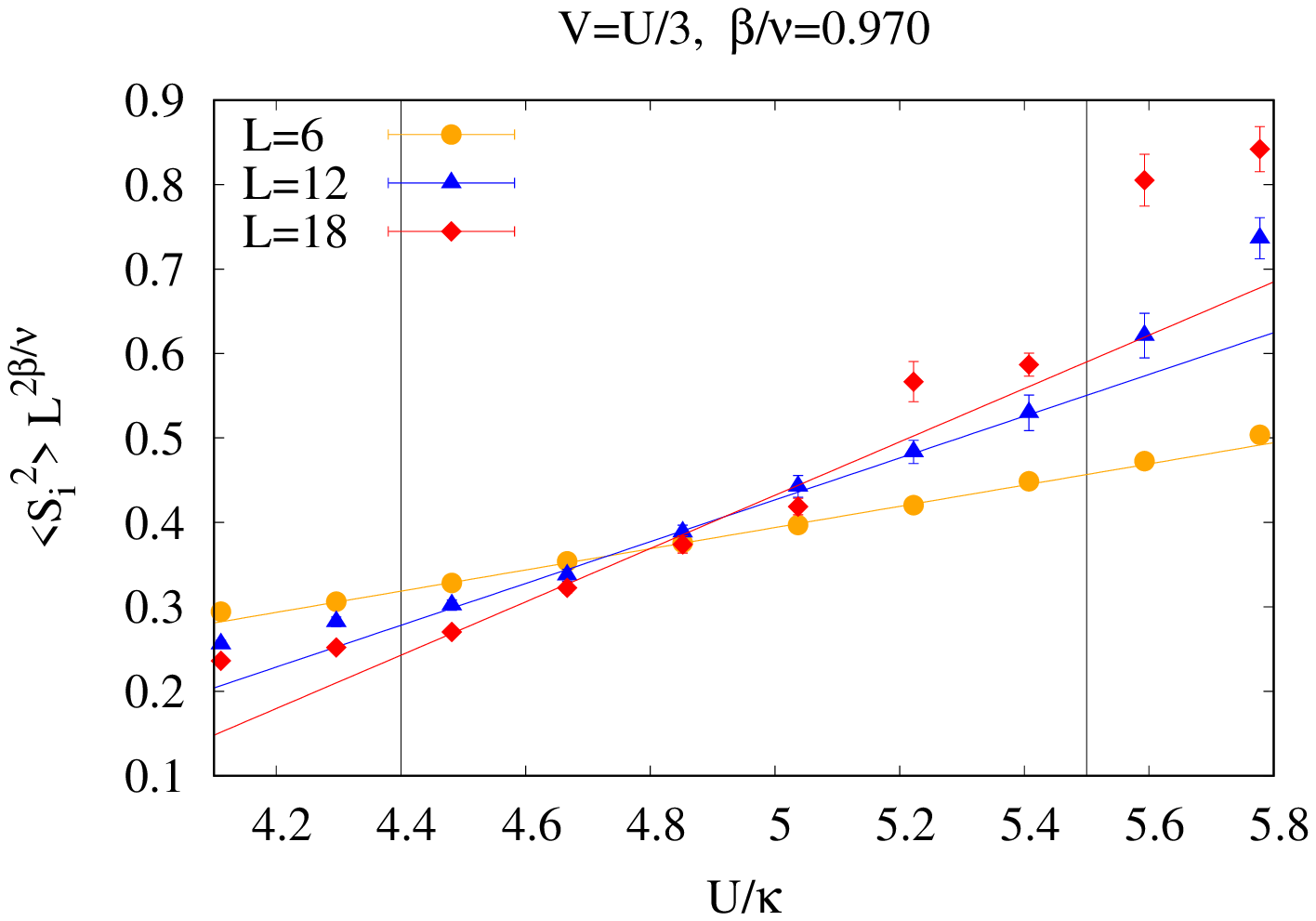}
\includegraphics[width=0.47\linewidth]{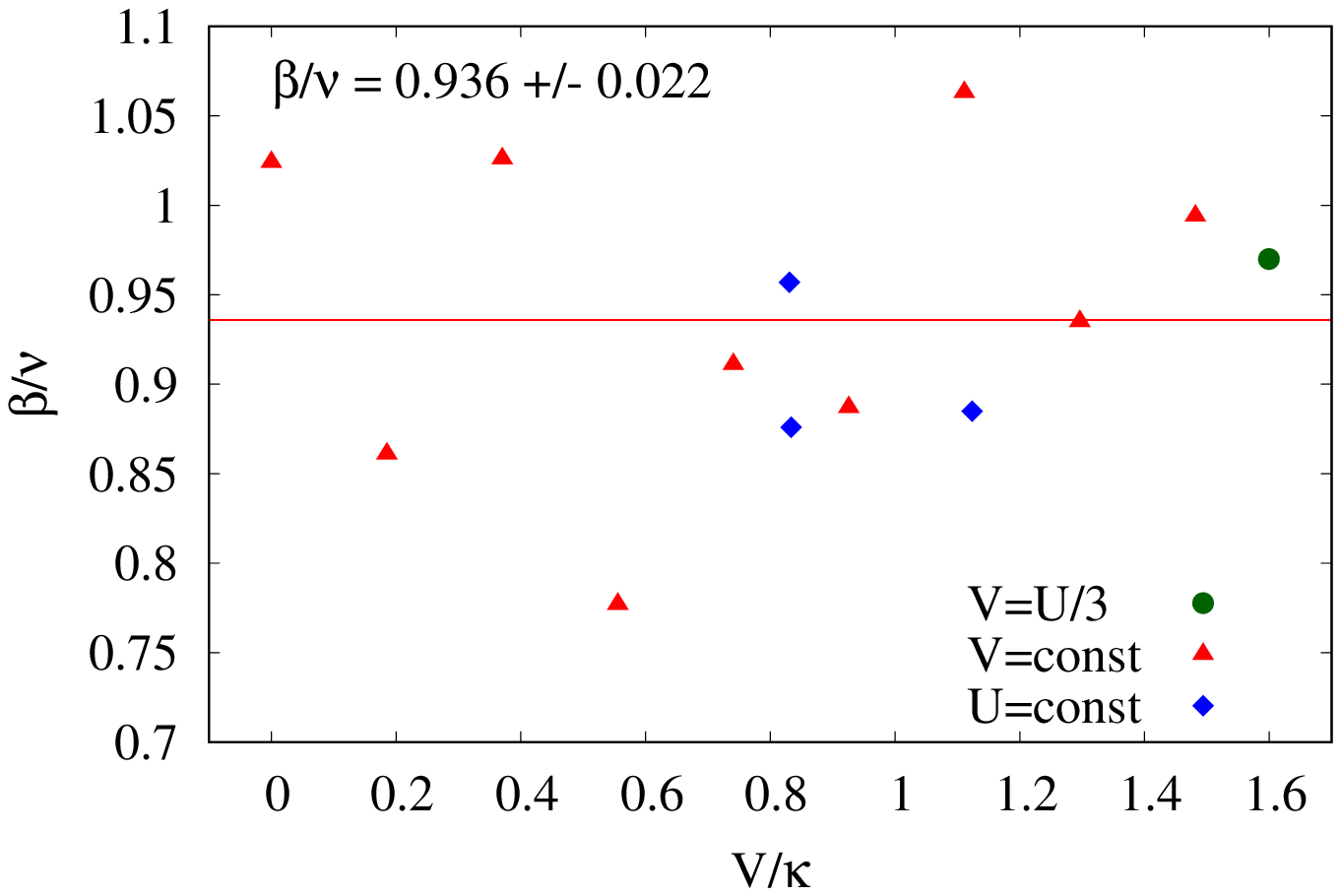}
\caption{Figures taken from \cite{Buividovich:2018hubb}. \emph{Left:} Finite size scaling of $\langle S^2 \rangle$ for $V_1=U/3$ (example).
$\beta/\nu$ is chosen such that the enclosed triangle
between fits to $L=6,12,18$ is minimized. \emph{Right:} Optimal exponents $\beta/\nu$ for horizontal, vertical
and diagonal lines in the phase diagram.}
\label{fig:hubb_scaling}
\end{figure}

The requirement of a positive definite interaction matrix leads to the restriction $V_1<U/3$.
To obtain the phase diagram in this region, we used a two step process. First we carried out infinite-volume extrapolations of
$\sqrt{\langle S^2 \rangle}$ and $\sqrt{\langle q^2 \rangle}$. Both of these quantities turned out to
have an approximately linear dependence on $1/L$, so the intercept of fits of the form $a+b\cdot(1/L)$ could
be used to estimate the extent of ordered phases. This revealed that the SDW phase at large $U$ extends
all the way to the $V_1=U/3$ line, while no evidence for CDW order was found. To supplement this result,
a finite size scaling analysis of $\langle S^2 \rangle$ was carried out close to the presumed phase
boundary, along horizontal, vertical and diagonal lines. This revealed that the second order phase transition
of the pure on-site model extends to $V_1=U/3$, and is likely described by the same critical exponents
$\beta/\nu\approx 0.9$, consistent with Gross-Neveu universality. The results are
summarized in figures \ref{fig:hubb_phase} and \ref{fig:hubb_scaling}, which are taken from
\cite{Buividovich:2018hubb}.

\section{Gap formation in graphene}
\label{sec:graphenephase}

While it is now firmly established that graphene in vacuum is a conductor, the question of the closest gapped phase in its phase diagram remains of interest, in particular in light
of the rapid development of experimental techniques to control the microscopic interaction parameters
and the emergence of other hexagonal materials. A crucial question thereby is the role of the Coulomb tail
of the two-body potential, which is essentially unscreened in $2D$ materials. Strong-coupling expansions
and renormalization group studies have suggested that dynamical gap formation is driven primarily by
short-range interactions, which would imply a semimetal-SDW transition with Gross-Neveu universality for
sufficiently strong coupling \cite{Semenoff_2012,PhysRevLett.97.146401}.
In contrast, a Dyson-Schwinger study of the low-energy effective field theory of graphene, which is
sensitive only to the long-range physics, predicted a (pseudo) conformal phase transition (CPT)
governed by exponential (``Miransky'') scaling \cite{Gamayun:2010}.
In principle the long-range part of the potential might also favor other phases, with CDW
being the most likely candidate.

\begin{figure}
\begin{center}
\includegraphics[width=0.43\linewidth]{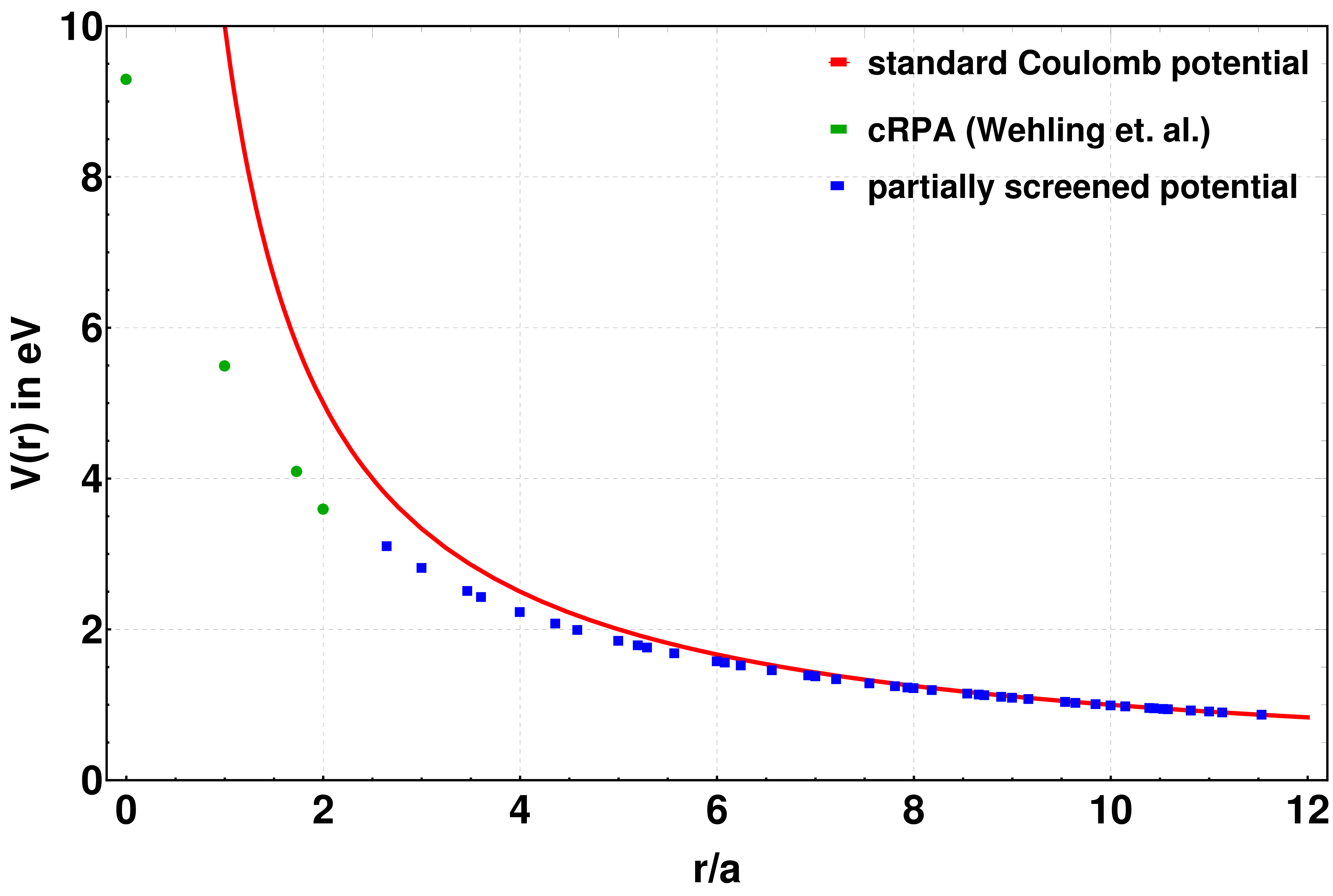}
\includegraphics[width=0.50\linewidth, trim=0 0.55cm 0 0]{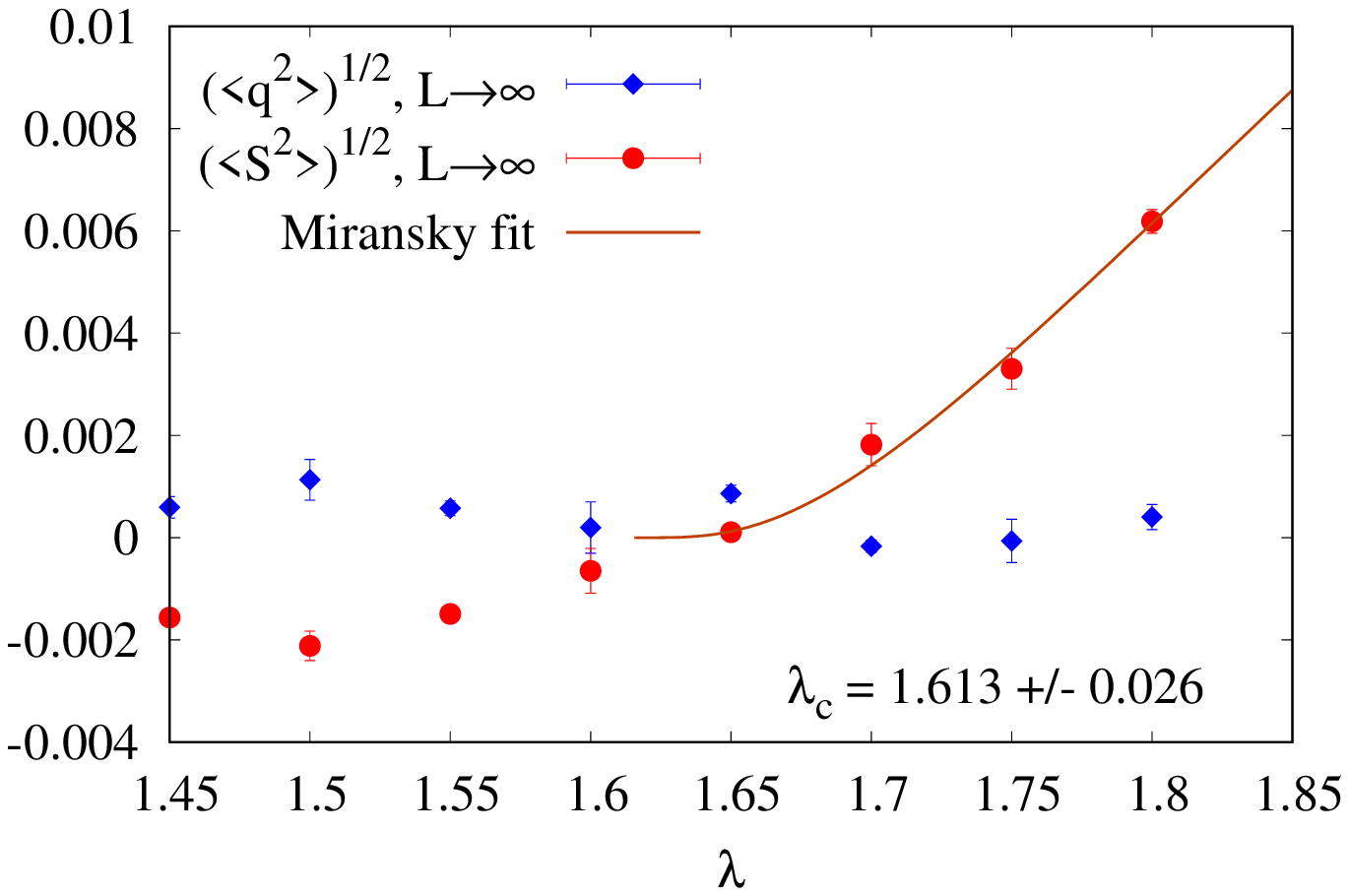}
\caption{Figures taken from \cite{Buividovich:2018crq}. \emph{Left:}
Partially screened Coulomb potential compared with unscreened Coulomb potential (red).
Points correspond to cRPA values (green) and thin-film model (blue) from \cite{PhysRevLett.106.236805}.
\emph{Right:} $L\to \infty$ limit of $\sqrt{\langle S^2 \rangle}$ (red) and $\sqrt{\langle q^2 \rangle}$ (blue). For spin a fit of (\ref{eq:miransky_scaling}) to points
at $\lambda>1.61$ is shown.
}
\label{fig:miransky1}
\end{center}
\end{figure}

In the past we studied gap formation in graphene intensively using HMC \cite{Buividovich:2012nx,Ulybyshev:2013swa,Smith:2014tha}.
These early works were focused on the detection of SDW order however, and employed source terms
which explicitly favor such a phase. Moreover, the critical properties were never addressed.
More recently, we thus decided to revisit this topic in a fully unbiased study without explicit sources
\cite{Buividovich:2018crq}, which also makes use of all of the most recent algorithmic developments,
in particular the complexification of the Hubbard field and the exponential action. We review the results
of this study here.
For $V_{xy}$ we used a realistic two-body potential for graphene, which contains the exact interaction parameters obtained
at short distances within a constrained random-phase approximation (cRPA) in \cite{PhysRevLett.106.236805}, and
is smoothly interpolated to an unscreened Coulomb tail using a thin-film model (see figure \ref{fig:miransky1}, left).
To drive the system towards the semimetal-insulator phase transition, we rescaled this potential by a factor $\lambda > 1$,
so that suspended graphene corresponds to $\lambda=1$.

We considered the same observables as in section \ref{sec:hubbphase},
i.e. the squared spin $\langle S^2 \rangle$ and squared charge $\langle q^2 \rangle$ per sublattice as defined by
(\ref{eq:squarespin}), or their square roots respectively.
Using linear fits of the form $f(1/L)=a+b\cdot(1/L)$ to lattice sizes $L=12,18,24$ we carried out an $L\to \infty$
extrapolation of $\sqrt{\langle S^2 \rangle}$ and $\sqrt{\langle q^2 \rangle}$ for a range of $\lambda$ values
(see figure \ref{fig:miransky1}, right panel). We can immediately conclude that SDW order is favored over CDW order: while the
extrapolation of $\sqrt{\langle q^2 \rangle}$ is consistent with zero
for any of the coupling strengths considered, $\sqrt{\langle S^2 \rangle}$ develops a non-zero expectation value around $\lambda_c \approx 1.65$.
By fitting the Miransky scaling function
\begin{equation}
\sigma(\lambda)=a\exp\left(\frac{-b} {\sqrt{\lambda - \lambda_c}} \right),
\label{eq:miransky_scaling}
\end{equation}
as appropriate for reduced QED$_4$ (see \cite{Gamayun:2010}) to $\sqrt{\langle S^2 \rangle}|_{L=\infty}$, we can
estimate $\lambda_c=1.61 \pm 0.02$. Note that the quality of this fit does not necessarily imply the validity of the CPT
scenario. A power-law fit works equally well, which reflects the difficulty of detecting Miransky scaling with direct methods.

\begin{figure}
\begin{center}
\includegraphics[width=0.47\linewidth]{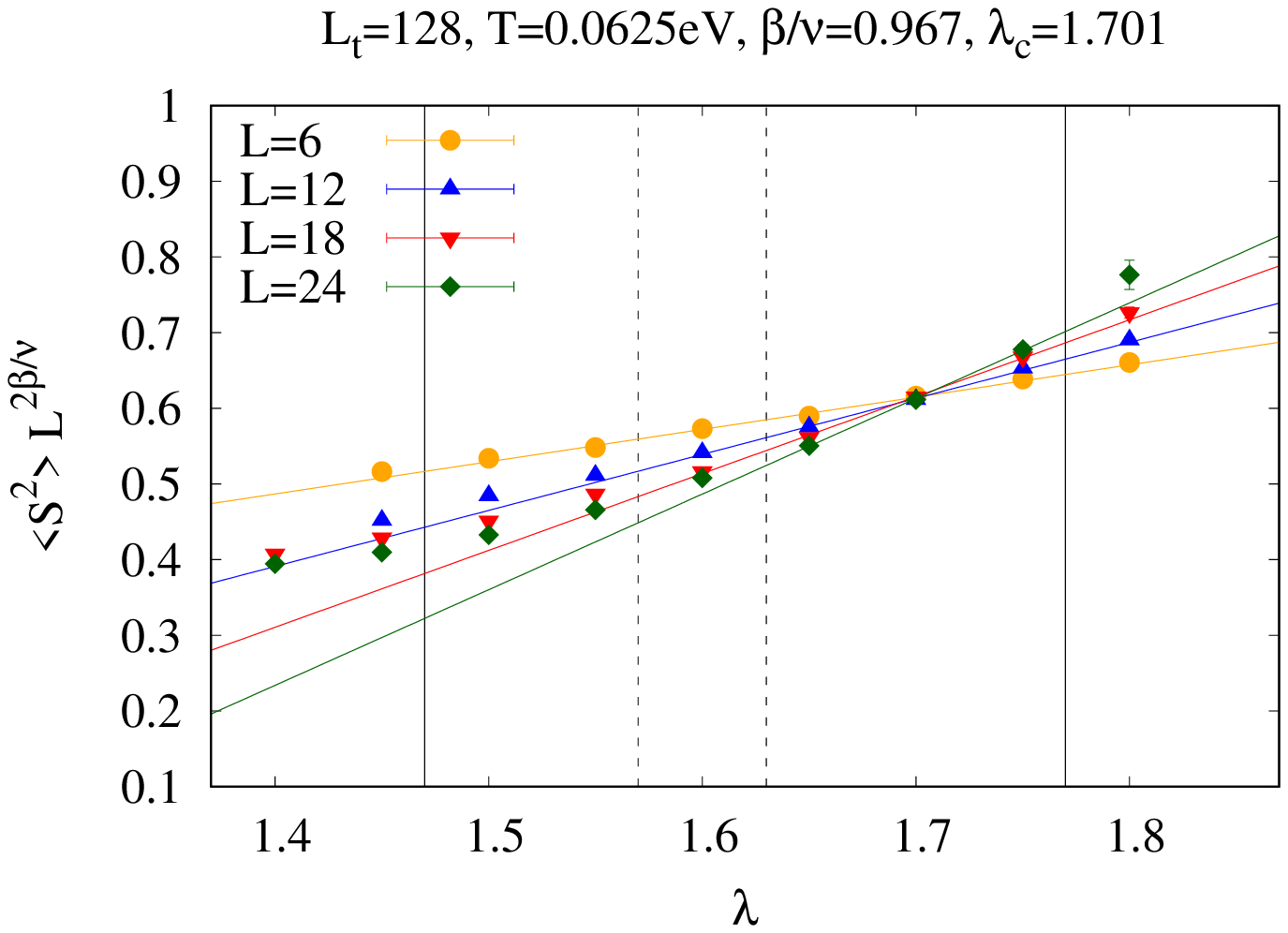}
\includegraphics[width=0.47\linewidth]{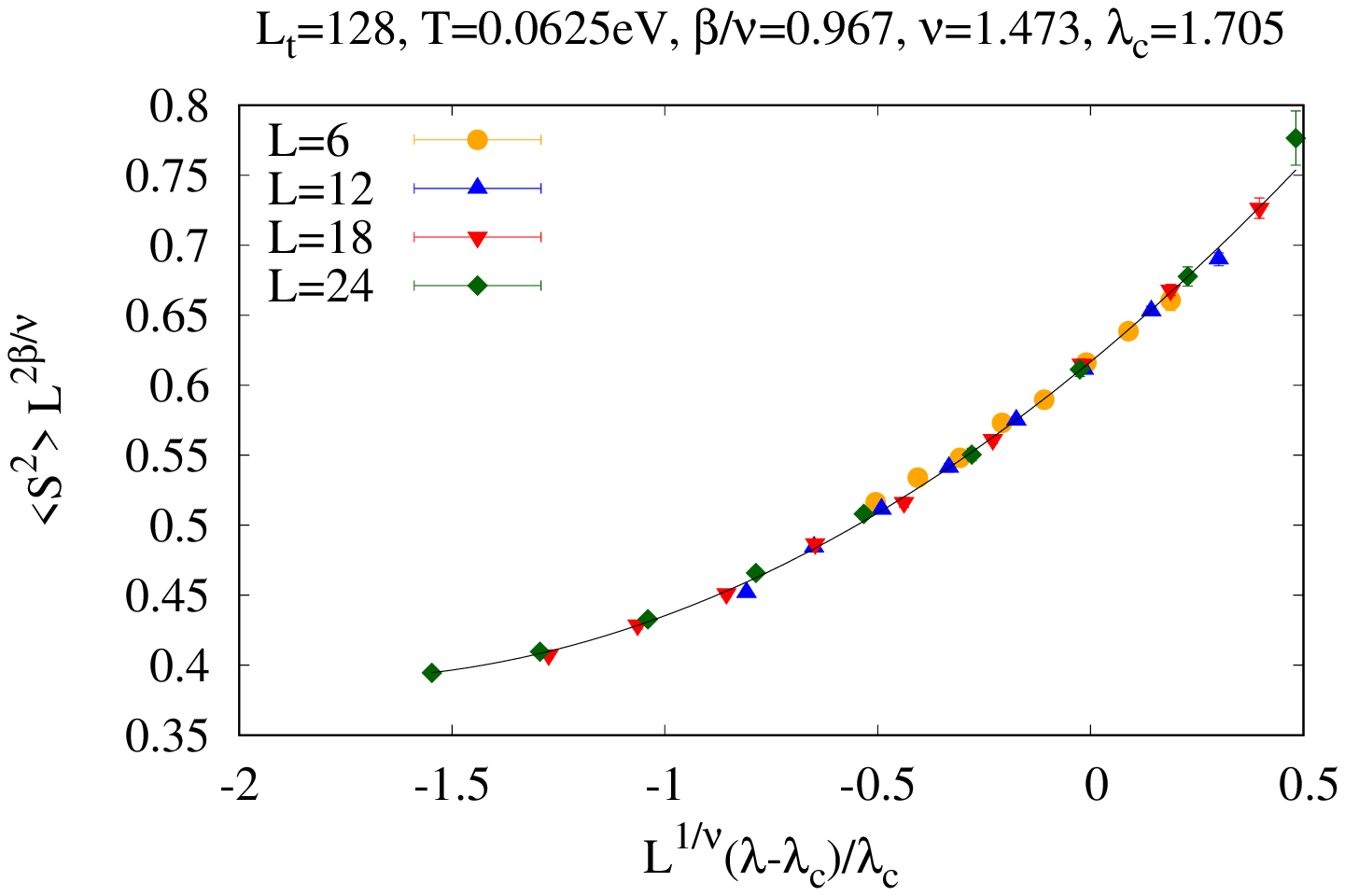}
\caption{Figures taken from \cite{Buividovich:2018crq}. \emph{Left:} Finite size scaling of $\langle S^2 \rangle$ for graphene.
$\beta/\nu$ chosen to minimize enclosed triangles between linear fits to $L=6,12,18,24$.
Solid lines mark boundaries for all fits. Dashed lines mark lower bounds for $L=18,24$.
\emph{Right:} Collapse of $\langle S^2 \rangle L^{2\beta/\nu}$ onto universal scaling function.
$\beta/\nu$ taken from left panel. $\nu$ and $\lambda_c$ were optimized by minimizing the
    $\chi^2/\textrm{d.o.f.}$ of fits of all data to polynomial functions of $x=L^{1/\nu}(\lambda-\lambda_c)/\lambda_c$.}
\label{fig:miransky2}
\end{center}
\end{figure}

To determine the critical properties, we first carried out a finite-size scaling study of $\langle S^2 \rangle$.
This was done by applying linear fits to $\langle S^2 \rangle L^{2\beta/\nu}$ as a function of $\lambda$, and determining
$\beta/\nu$ by minimizing the enclosed triangles between fits to $L=6,12,18,24$. While this procedure leads to
$\beta/\nu \approx 0.97$ (slightly larger than the estimates for Gross-Neveu universality) for a particular choice of fit windows
(see left panel of figure \ref{fig:miransky2}), we find that $\beta/\nu$ is not constrained very strongly by our data: by choosing
different fit windows it is possible to obtain values in the range $0.95\ldots 1.0$, with corresponding estimates for $\lambda_c$
in the range $1.6\ldots 1.7$. We stress that this is rather unusual for a second order phase transition. In fact, we applied
the same procedure to a set of test data obtained with onsite interactions $U$ only, and found that both
$\beta/\nu$ and $U_c$ are tightly constrained in that case.

\begin{figure}
\begin{center}
\includegraphics[width=0.47\linewidth]{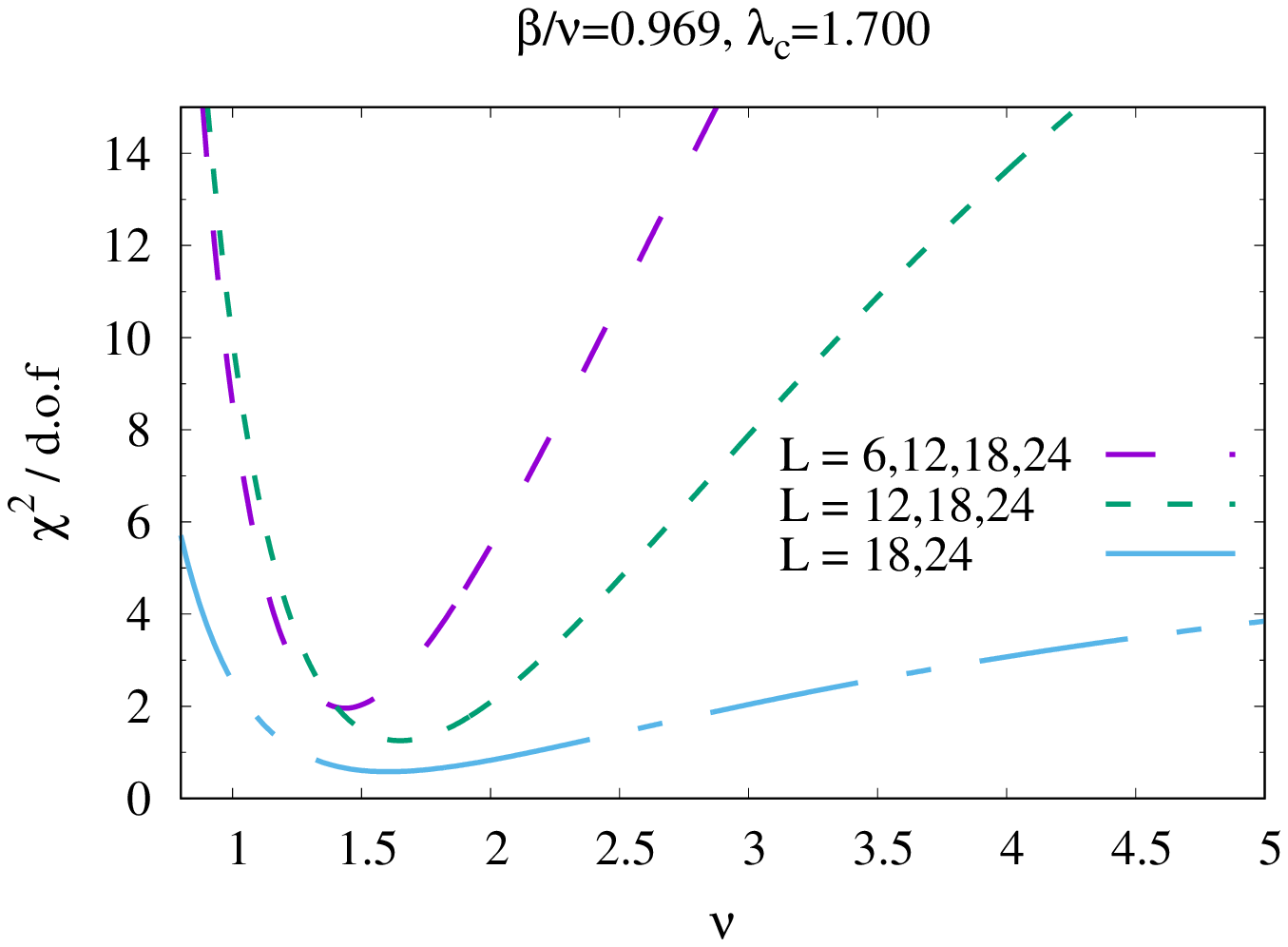}
\includegraphics[width=0.47\linewidth]{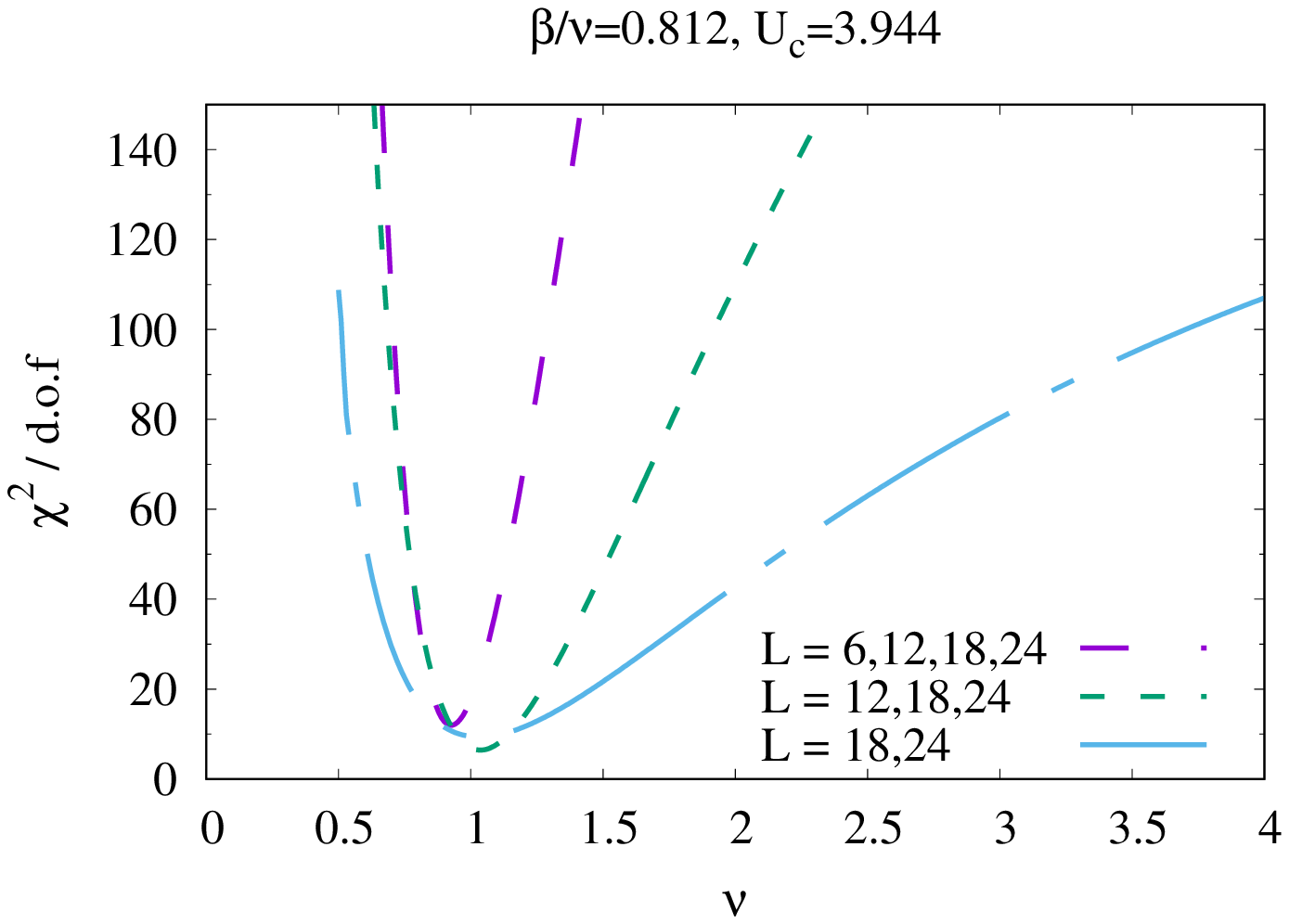}
\caption{$\chi^2/\textrm{d.o.f}$ for collapse of $\langle S^2 \rangle L^{2\beta/\nu}$ onto a universal finite-size scaling function for
graphene (left) and Hubbard model (right) with different choices of $\nu$ for different sets of lattice sizes. $\beta/\nu$ and critical
coupling strength are fixed to their optimal values. For graphene the quality of fit is rather insensitive to increases of $\nu$ when
smaller lattices are excluded. Figures taken from \cite{Buividovich:2018crq}.}
\label{fig:miransky3}
\end{center}
\end{figure}

We also attempted to collapse all data-points onto a univeral scaling function, by plotting  $\langle S^2 \rangle L^{2\beta/\nu}$
over $x=L^{1/\nu}(\lambda-\lambda_c)/\lambda_c$ and choosing $\nu$ such that the
$\chi^2/\textrm{d.o.f.}$ of a fit of all data to a polynomial function $f(x)$ is minimized. This reveals a similarly puzzling
situation. First, we find that the optimal exponent $\nu\approx 1.47$ is again much larger than for the Gross-Neveu model (which
is $\nu\approx 1.2$). Perhaps more importantly however, we find that the constraints placed on $\nu$ again are very weak,
with the quality of fit being not very sensitive to increases of $\nu$. This effect becomes even more pronounced when smaller lattice
sizes are ignored. If we exclude all but the $L=18,24$ data from the fit we find that it is possible to increase $\nu$ almost
indefinitely. This is very different from the pure Hubbard model, where we made an unsuccessful attempt to reproduce this feature.
See figure \ref{fig:miransky3} for an illustration.

We propose that the CPT offers a natural explanation for our results.
In \cite{Gamayun:2010} it was argued that the CPT formally corresponds to the limit $\beta, \nu\to\infty$, $\delta=1$ of a
second order transition and that the usual hyperscaling relations may apply. In our case $d=2$ (where $d$ is the number of
spatial dimensions) and with $\delta=1$ the relation
\begin{equation}
\frac{\beta}{\nu}=\frac{d}{\delta+1} \label{eq:hyperscaling}
\end{equation}
would thus lead to $\beta/\nu=1$, which agrees with our estimate at a level of about  $3 \%$ and is well within the present errors.
A CPT is also expected to receive power-law corrections and mimic a second order transition in finite volume. It is thus reasonable
to assume that $\beta, \nu\to\infty$ is obtained in the infinite volume limit, with their ratio fixed by (\ref{eq:hyperscaling})
on finite lattices.

\begin{figure}
\begin{center}
\includegraphics[width=0.47\linewidth]{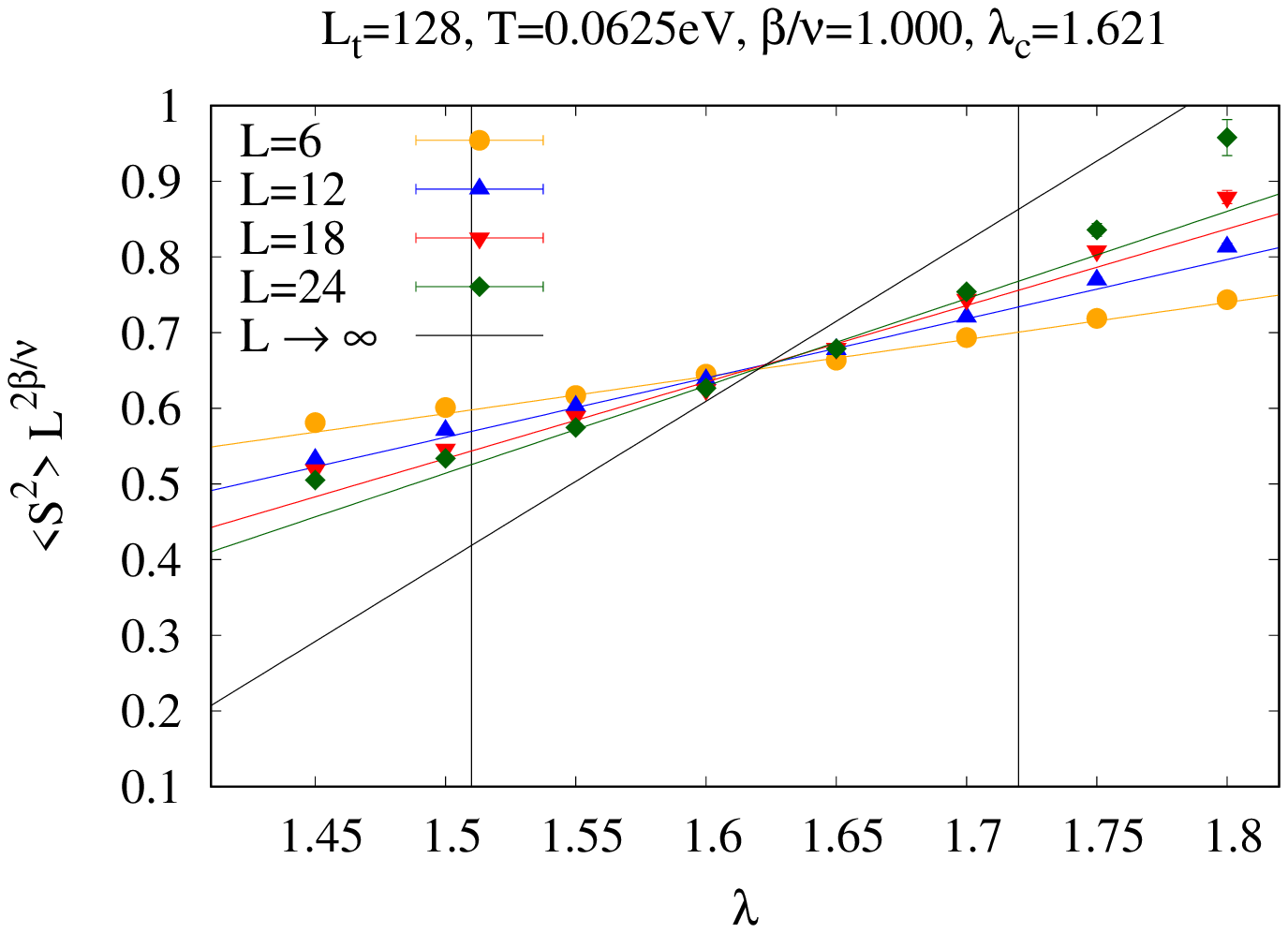}
\includegraphics[width=0.47\linewidth]{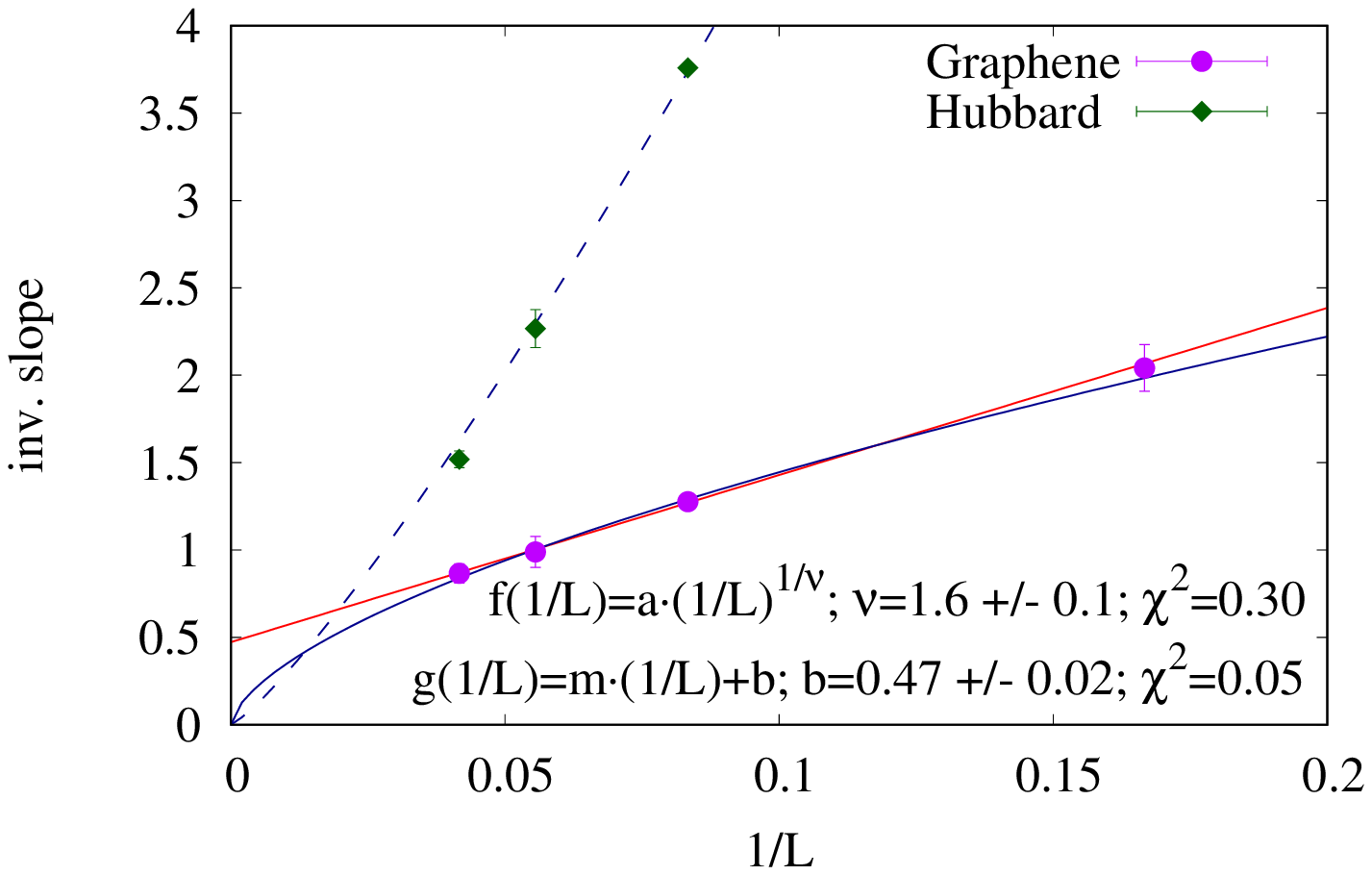}
\caption{Figures taken from \cite{Buividovich:2018crq}. \emph{Left:} Finite size scaling of $\langle S^2 \rangle$ for graphene with fixed $\beta/\nu=1.0$.  Linear fits
bounded by vertical lines. \emph{Right:} Extrapolation of the inverse slope of the fits to $L=\infty$
limit, using $g(1/L)=m\cdot(1/L)+b$ and $f(1/L)=a\cdot(1/L)^{1/\nu}$ respectively, where $\nu$ is a free parameter.
$g(\cdot)$ (with smaller $\chi^2$) predicts that the slope remains finite for $L\to \infty$, suggesting that the
data collapse onto a universal function without a rescaling of $\lambda$. Results for the Hubbard model with
a power-law model curve shown for comparison.}
\label{fig:miransky4}
\end{center}
\end{figure}

To test this we fixed $\beta/\nu=1$ and plotted $\langle S^2 \rangle L^{2\beta/\nu}$ over $\lambda$
(see figure \ref{fig:miransky4}, left panel). We find linear fits
to the data intersect at around $\lambda\approx1.62$, which is consistent within error with
$\lambda_c=1.61 \pm 0.02$ obtained from the fits to (\ref{eq:miransky_scaling}) discussed above.
More importantly, we note that the slopes in the intersection plot do not appear to increase towards infinity with
increasing volumes as they should for a second-order phase transition. This is consistent with $\nu\to \infty$,
as it would imply a collapse of all data without a rescaling of the coupling constant.
To investigate this somewhat more carefully, we plot the inverse slopes of our linear fits to $\langle S^2 \rangle L^{2\beta/\nu}$ from the intersection plots over $1/L$ in the right panel of figure \ref{fig:miransky4}:
While test data for the Hubbard model once again shows the expected behavior, the inverse slopes for graphene
(with $\beta/\nu=1.0$) are well described by a linear model fit to
\[
g(1/L)=m\cdot(1/L)+b \quad\mbox{with}\quad  b=0.47 \pm 0.02\, .
\]
This non-zero intercept $b$ then provides our best numerical evidence of a finite slope in the infinite-volume limit and hence of $\beta,\nu \to \infty$ as CPT characteristics.

\section{Lifshitz transition in graphene}
\label{sec:vhs}

\begin{figure}
\begin{center}
\includegraphics[width=0.51\linewidth, trim = 0 -3cm 0 0]{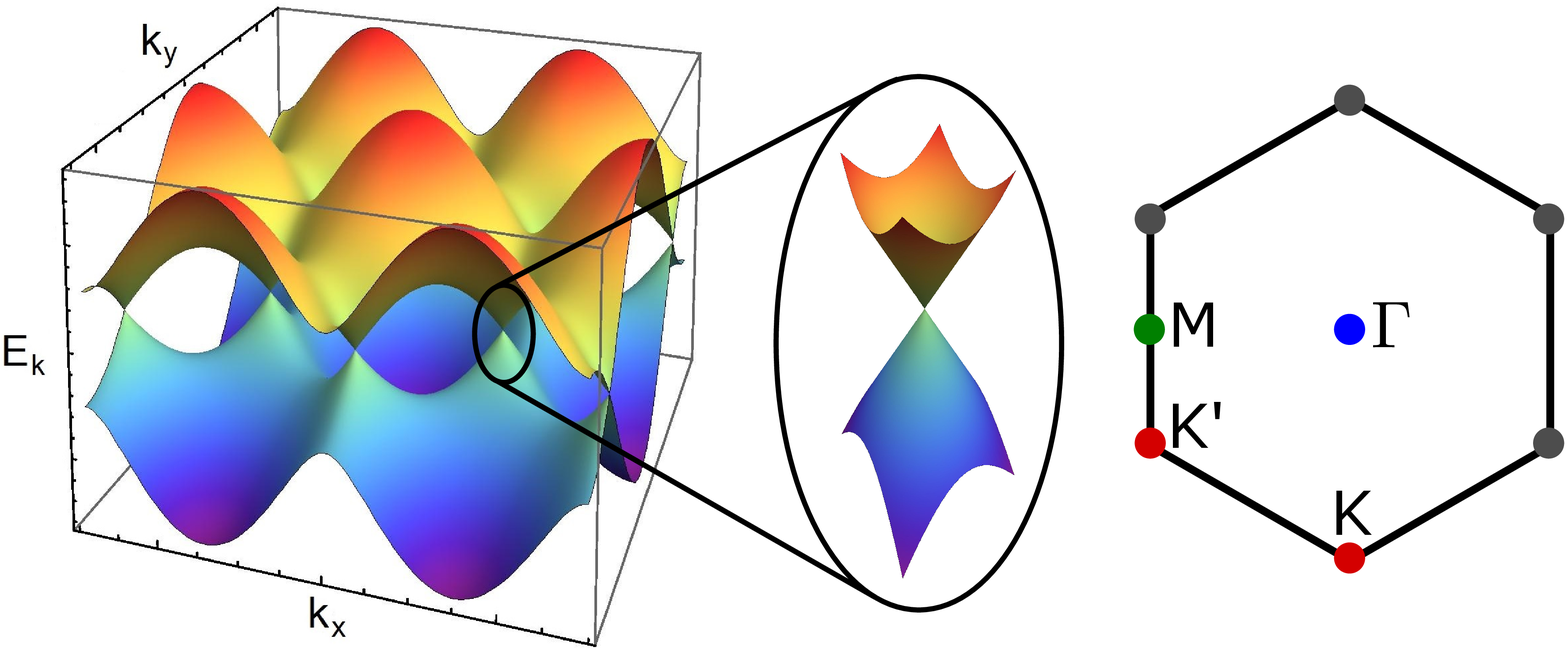}
\includegraphics[width=0.44\linewidth]{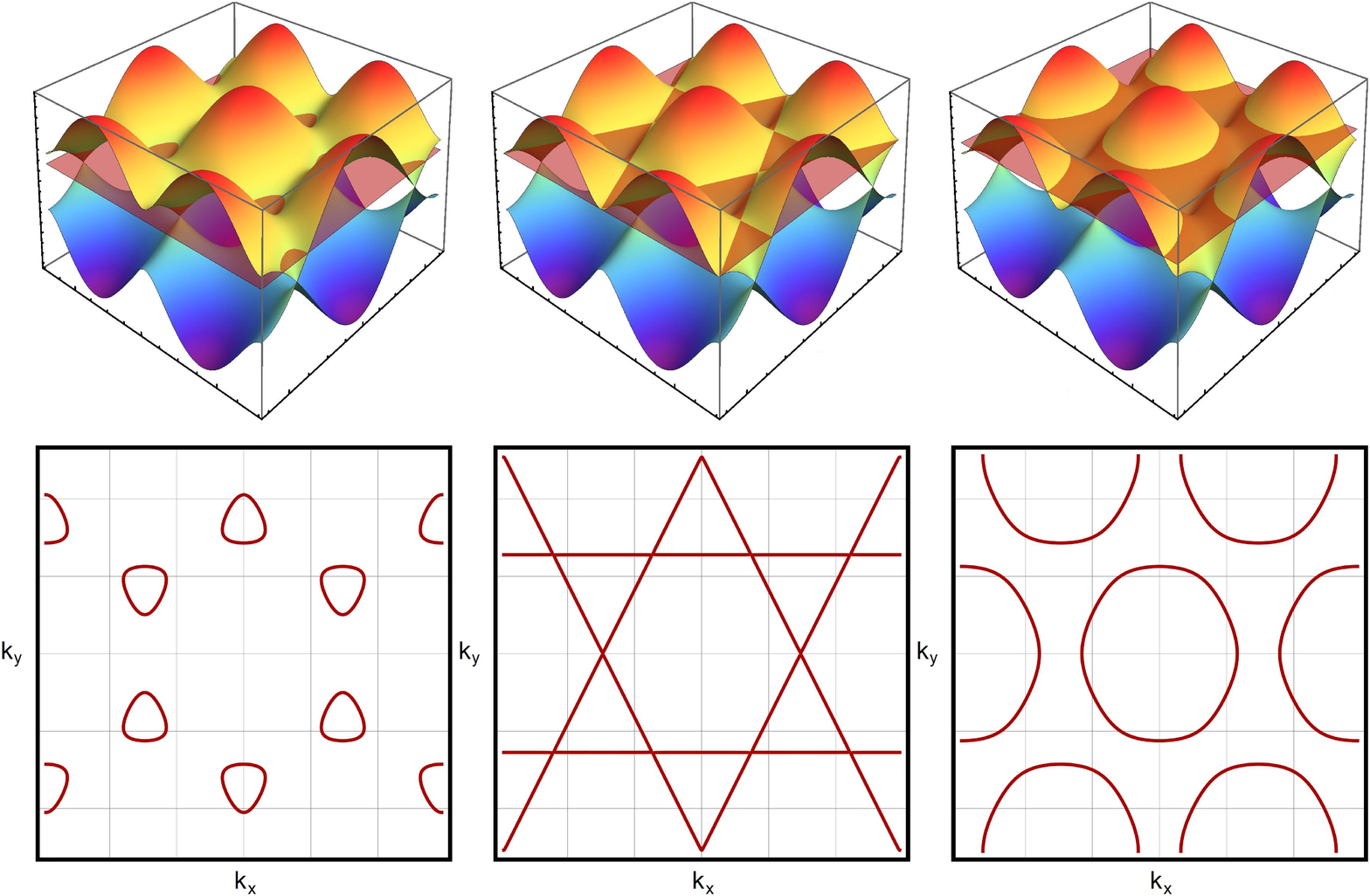}
\caption{Figures taken from \cite{Korner:2017qhf}. \emph{Left:} Band structure of tight-binding
theory of graphene. Dirac cones around the K-points are enlarged. The first Brillouin zone
and terminology for special points are shown.
\emph{Right:} Topology of the Fermi lines (intersection lines with horizontal planes) for Fermi
levels below (left), exactly at (middle) and above (right) the saddle points. }
\label{fig:lifshitz1}
\end{center}
\end{figure}

In the electronic bands of the non-interacting tight-binding theory one finds saddle points,
located at the M-points, which are characterized by a vanishing group velocity
(see left panel of figure \ref{fig:lifshitz1} for an illustration of the band structure).
These separate the low-energy region, described by an effective Dirac
theory, from a region where electronic quasi-particles behave like a regular Fermi liquid with a parabolic
dispersion relation centered around the $\Gamma$-points. When the Fermi level is shifted across the saddle points by a
chemical potential $\mu$, a change of the topology of the Fermi surface (which is one-dimensional for a 2D crystal)
takes place. The distinct circular Fermi (isofrequency) lines surrounding the Dirac points are deformed into triangles,
meet to form one large connected region and then break up again into circles around the $\Gamma$-points
(see figure \ref{fig:lifshitz1}, right). This is known as neck-disrupting Lifshitz transition (NDLT) and
occurs exactly at $\mu=\kappa$ (where $\kappa$ is the hopping energy) in the non-interacting theory.
It also leads to a logarithmic divergence of the peak height of the density-of-states (DOS)
with increasing surface area of the graphene sheet, known as a van Hove singularity (VHS).

The Lifshitz transition is not a true phase transition in the thermodynamic sense, as it is purely
topological and not associated with any type of symmetry breaking.
Since interactions are strongly enhanced by the divergent DOS, it is generally believed however that
the VHS is unstable towards formation of electronic ordered phases if many-body interactions are
accounted for. This would imply that the
Lifshitz transition becomes a true phase transition in a realistic description of the interacting
system at low temperatures. An exciting possibility is the emergence of an
anomalous time-reversal symmetry violating chiral d-wave superconducting phase in graphene \cite{ChubukovNature}.
In light of the recent discovery
of a tunable superconducting gap in twisted bilayer graphene, which exists only for certain ``magic'' twist angles
\cite{Cao2018UnconventionalSI}, there is also renewed interest in the VHS, as a possible driving-mechanism 
for this instability \cite{PhysRevB.98.205151}.

In \cite{Korner:2017qhf} we studied the fate of the VHS in the presence of interactions at finite spin density
using HMC. This was done by adding a Zeeman-splitting term to the Hamiltonian, which shifts the Fermi levels of the two
spin orientations in opposite directions.
In the non-interacting limit, finite spin and finite charge density are
indistinguishable, and either one may be used to characterize the NDLT.
Unlike finite charge density, simulations at finite spin density are not affected by a fermion sign problem
and allow us to probe genuine interaction effects on the VHS. Physically, our setup corresponds to an in-plane
magnetic field, which is interesting in its own right. Simulations were carried out using the linearized fermion matrix,
purely imaginary Hubbard fields, and a spin-staggered mass term of $m_s=0.5\, \textrm{eV}$ to remove zero modes.
We used the same partially screened Coulomb potential as discussed in section \ref{sec:graphenephase}, with
a rescaling parameter $\lambda$ to control the interaction strength.

The logarithmic divergence of the DOS is manifest only at $T=0$ and is inaccessible to our simulations.
Instead, we thus focused on the ferromagnetic susceptibility $\chi$, which is related to the DOS through
the polarization or Lindhard function and
can also be used to characterize the NDLT (see \cite{Korner:2017qhf} for a detailed discussion). $\chi(\mu)$ is given by
\eq{ \chi(\mu)  &= - \frac{1}{N_c} \,
  \left(\frac{d^2\Phi}{d\mu^2} \right) = \frac{1}{N_c \beta} \left[\frac{1}{Z} \frac{d^2Z}{d\mu^2}
  -\frac{1}{Z^2} \left(\frac{dZ}{d\mu}\right)^2 \right]~, }
where  $\Phi=-T \ln{Z}$ is the grand-canonical potential and $N_c=N^2$ is the
number of unit cells. From this we can obtain
$\chi = \chi_\mathrm{con} + \chi_\mathrm{dis} $, with
\begin{eqnarray}
\hspace{-5mm}
\chi_\mathrm{con}(\mu) &=& \frac{-2}{N_c \beta}
\left\langle  \textrm{ReTr} \left( M^{-1} \frac{dM}{d\mu}M^{-1} \frac{dM}{d\mu} \right) \right\rangle,~\\
\hspace{-5mm}
\chi_\mathrm{dis}(\mu) &=& \frac{4}{N_c \beta}\left\{ \left\langle \left[ \textrm{ReTr}\left( M^{-1} \frac{dM}{d\mu} \right) \right]^2
\right\rangle - \left\langle \textrm{ReTr}\left( M^{-1} \frac{dM}{d\mu} \right) \right\rangle^2 \right\},
\label{eq:susc1}
\end{eqnarray}
where $\chi_\mathrm{con/dis}$ denote the connected and disconnected contributions respectively.
At any non-zero temperature, the infinite volume limit of $\chi(\mu)$
is bounded from above by some value $\chi_\mathrm{max}$, which diverges as the temperature
is lowered. In the non-interacting theory, this divergence occurs exactly at $\mu=\kappa$ and is described by
\begin{eqnarray}
\chi_\mathrm{max}(T,\mu=\kappa)  = -\frac{3g_\sigma }{2\pi^2\kappa}
   \ln{\big({\pi T}/{\kappa} \big)} + \gamma_E  + 3 \ln
2 + \mathcal{O}(T)~,
\label{eq:chimax}
\end{eqnarray}
where $\gamma_E$ is the Euler-Mascheroni constant and $g_\sigma=2$ for the spin degeneracy.
In this limit $\chi_\mathrm{dis}$ vanishes exactly since the expectation value $\langle \mathrm{ReTr}(\ldots)^2\rangle$
factorizes, and hence the divergence is driven purely by the connected part.

\begin{figure}
\begin{center}
\includegraphics[width=0.47\linewidth]{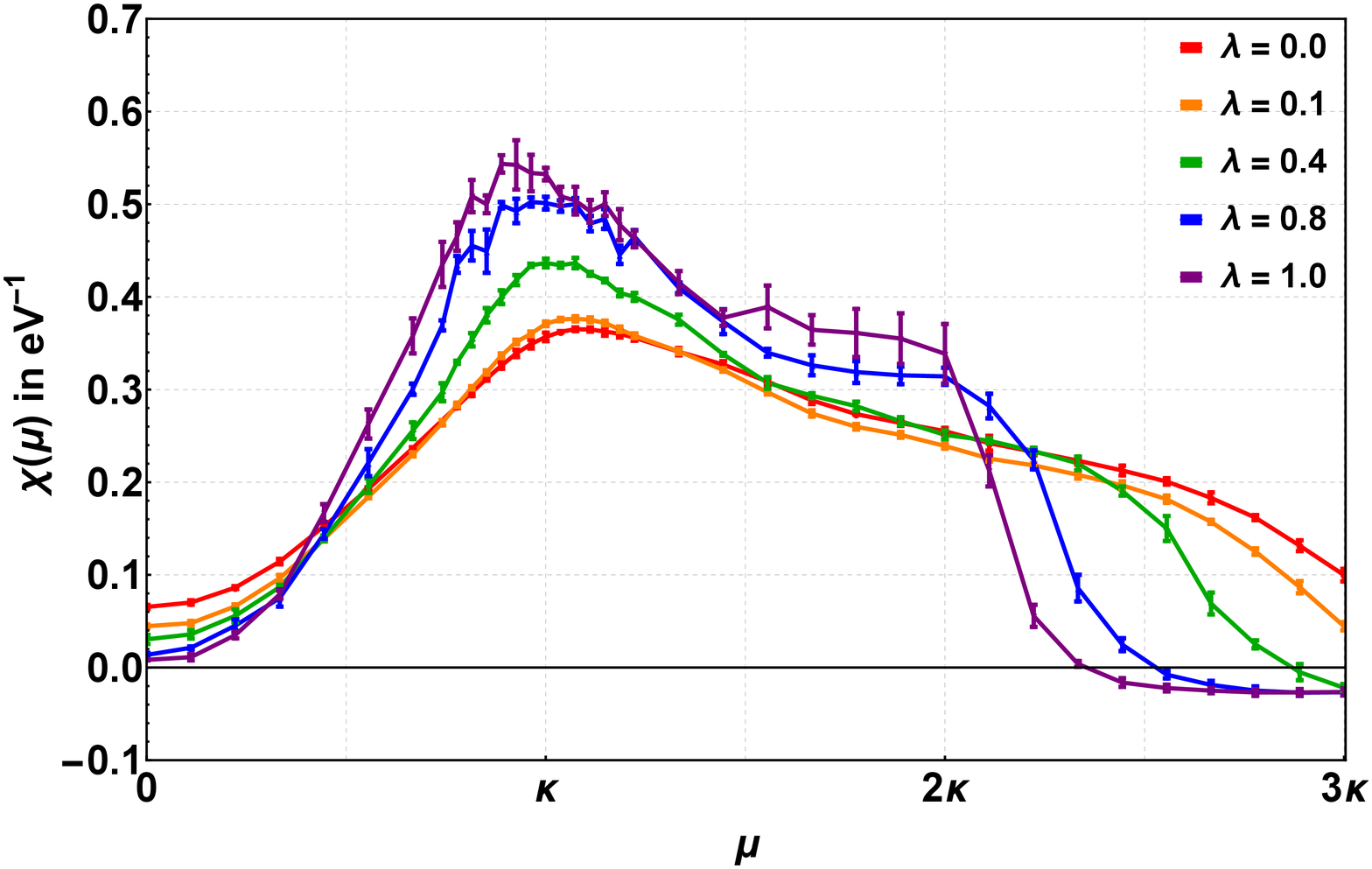}
\includegraphics[width=0.47\linewidth]{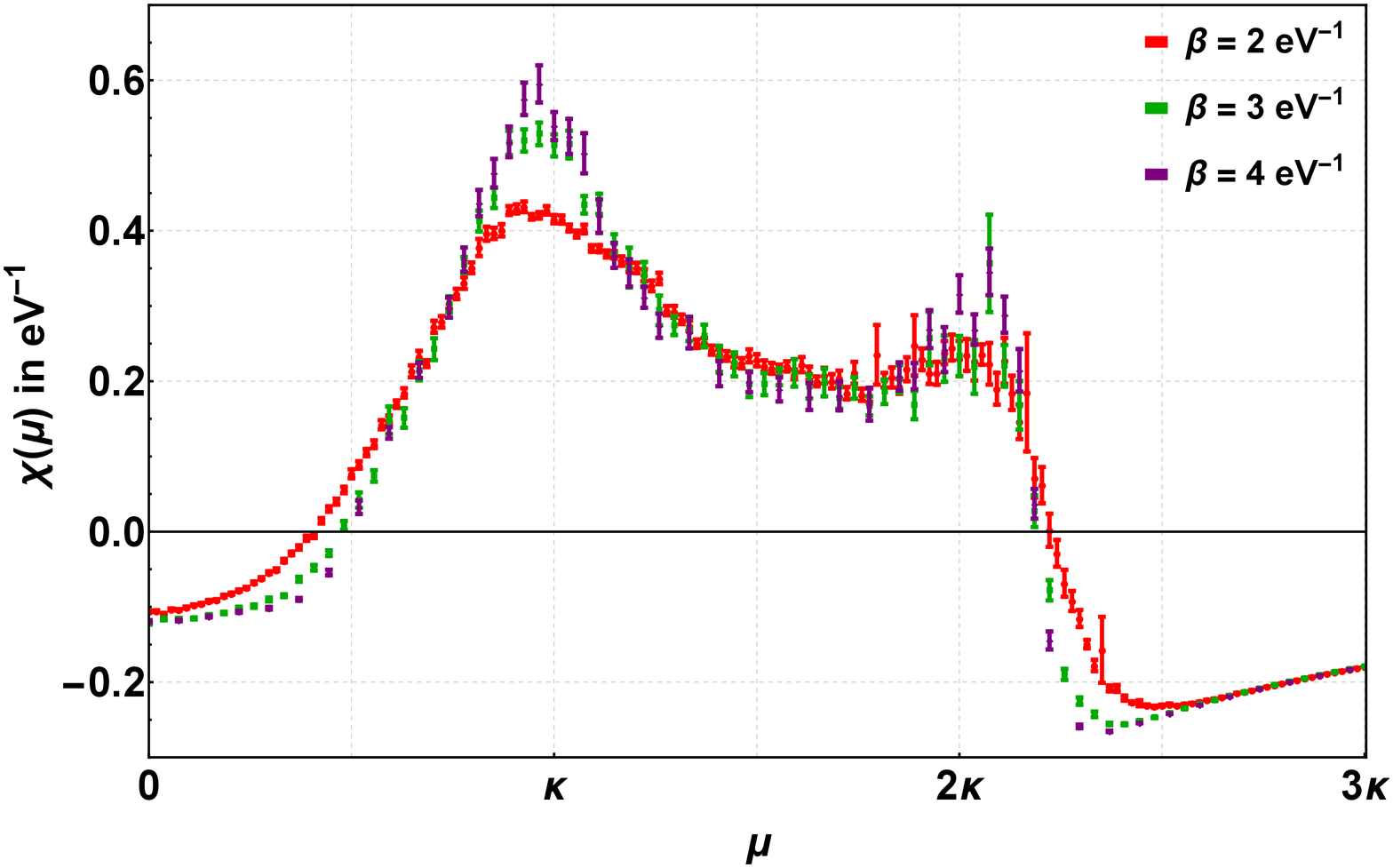}
\caption{Figures taken from \cite{Korner:2017qhf}. \emph{Left:}
$\chi(\mu)$ for $\beta=2\textrm{ eV}^{-1}$, $N = 12$ at different interaction strengths.
All points are quadratic $\delta_\tau \rightarrow 0$ extrapolations
from simulations at non-zero $\delta_\tau$.
 \emph{Right:}
Temperature dependence of $\chi(\mu)$. Lattice sizes scale linearly with $\beta$, such that the
displayed curves correspond to $N=12,18,24$ respectively;
with $\delta_\tau=1/6~ \textrm{eV}^{-1}$ and $\lambda=1$ for all cases.
}
\label{fig:lifshitz2}
\end{center}
\end{figure}

In figure \ref{fig:lifshitz2} (left) we show $\chi(\mu)$ with $N = 12$ at different
interaction strengths, ranging from the non-interacting theory to suspended graphene. These calculations
were done at fixed temperature with $\beta=2\textrm{ eV}^{-1}$ (to set the energy scale we chose
a hopping parameter of $\kappa = 2.7 \textrm{ eV}$ here, which roughly corresponds to the experimental
value of graphene). Each point was obtained
from simulations at several values of $\delta_\tau$ and extrapolated to the limit
$\delta_\tau \rightarrow 0$ using quadratic polynomials. We observe that the peak at the VHS
becomes more and more pronounced with increasing interaction strength. This is due
to both, a corresponding rise in the connected part at the VHS and an additional contribution from
the disconnected part. Likewise, the peak position as well as the upper end of
the conduction band are shifted towards smaller values of $\mu$.
Our results are in qualitative agreement with experimental data from
angle resolved photoemission spectroscopy (ARPES) measurements on charge-doped graphene
systems, which show evidence for a warping of the Fermi surface, leading to an extended,
not pointlike, van Hove singularity characterized by the flatness of the bands
along one direction, and hence a stronger divergence of the DOS than in the non-interacting
system \cite{PhysRevLett.104.136803}. Such experiments also observed bandwidth renormalization
(narrowing of the widths of the $\pi$-bands), due to interactions and doping \cite{PhysRevB.94.081403}.

\begin{figure}
\begin{center}
\includegraphics[width=0.47\linewidth]{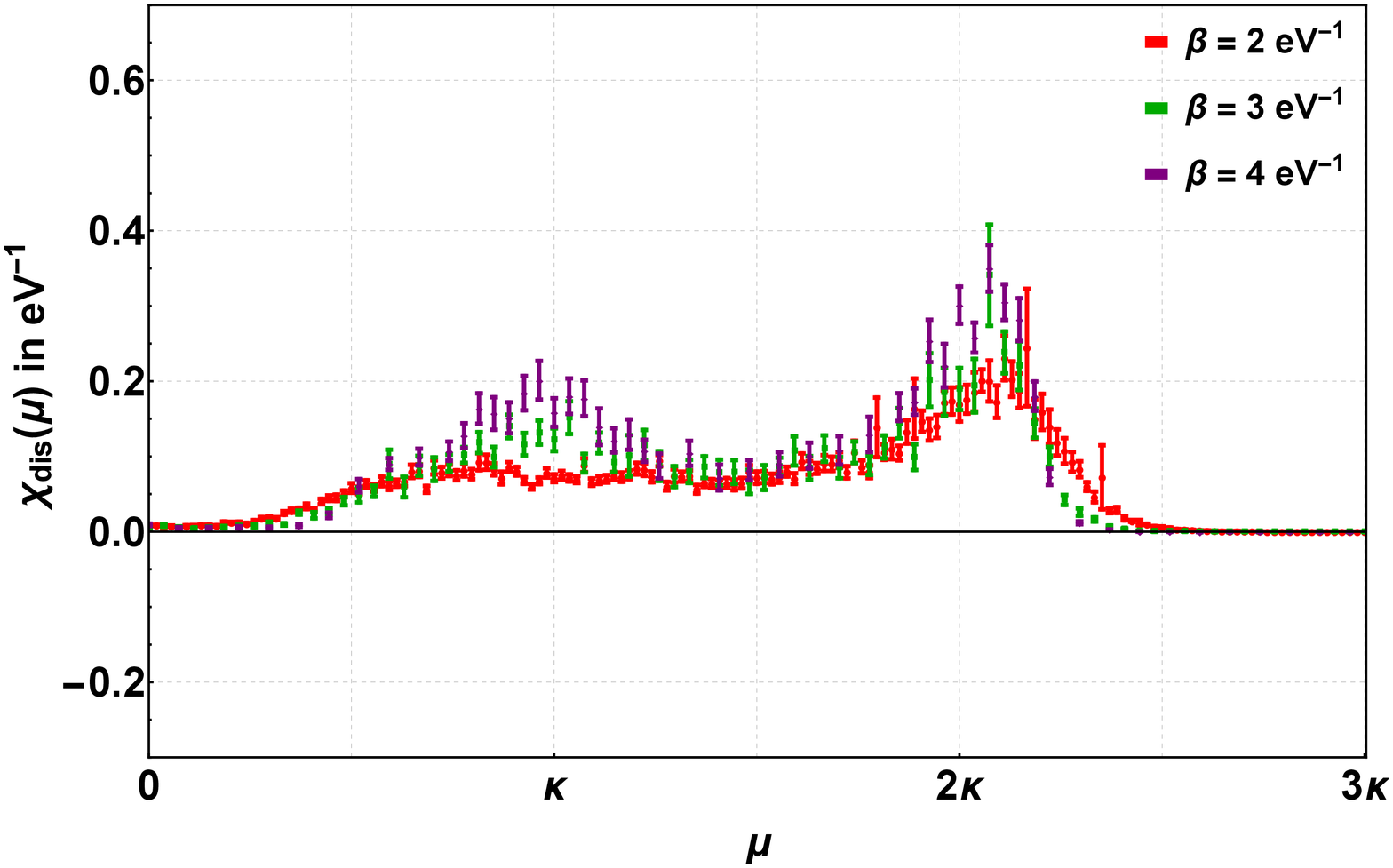}
\includegraphics[width=0.47\linewidth]{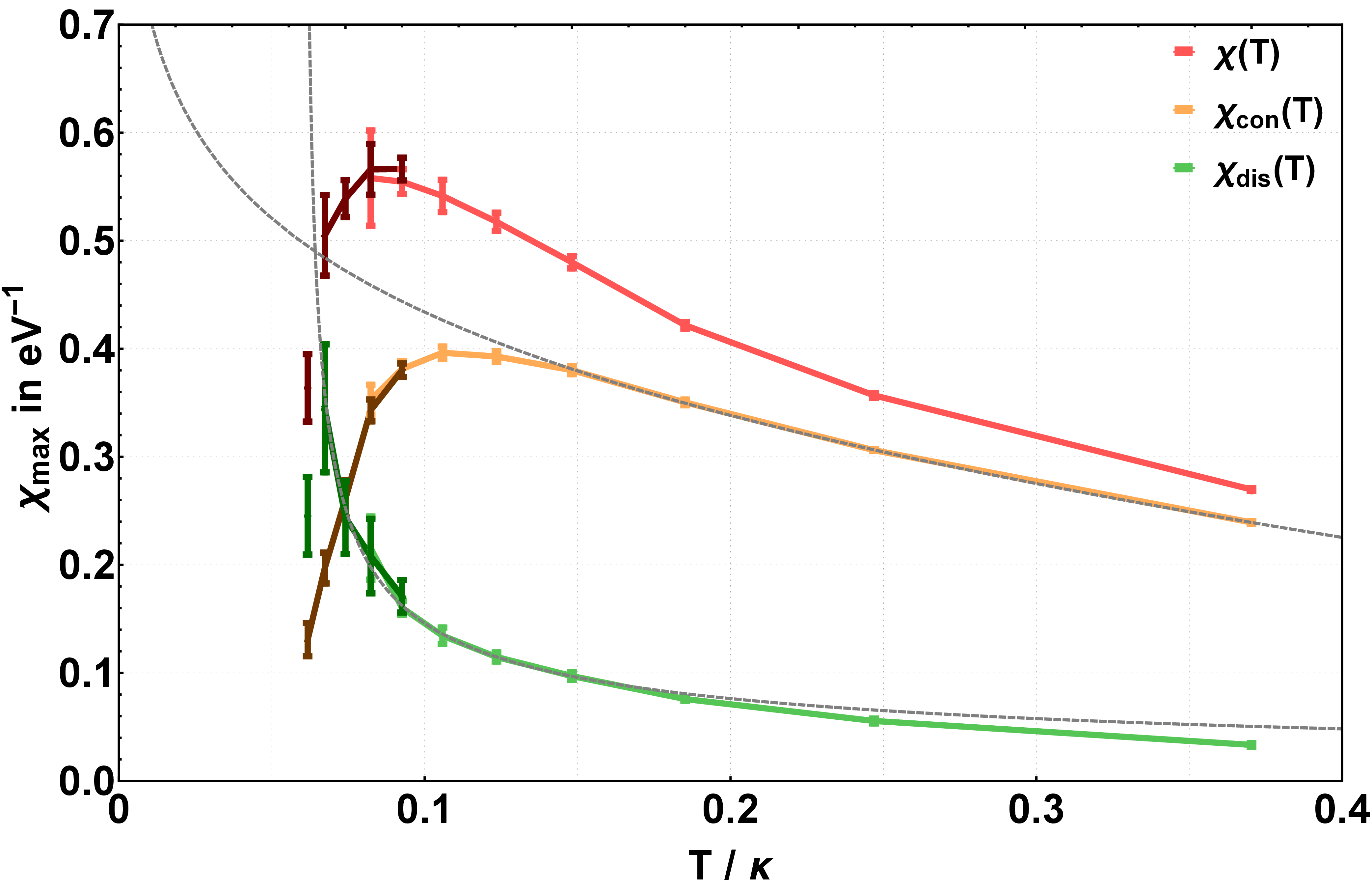}
\caption{Figures taken from \cite{Korner:2017qhf}. \emph{Left:} Temperature dependence of $\chi_\mathrm{dis}(\mu)$.
Lattice sizes scale linearly with $\beta$, such that the
displayed curves correspond to $N=12,18,24$ respectively;
with $\delta_\tau=1/6~ \textrm{eV}^{-1}$ and $\lambda=1$ for all cases.
\emph{Right:}
Temperature dependence of $\chi_\mathrm{max}$ in the range
$\beta=1.0, \ldots 6.0 \textrm{ eV}^{-1}$. Dotted lines
are fits using (\ref{eq:logdivergence}) for
$\chi_\mathrm{con}^\mathrm{max}$ and (\ref{eq:powdivergence}) for
$\chi_\mathrm{dis}^\mathrm{max}$ in appropriate ranges (see text).}
\label{fig:lifshitz3}
\end{center}
\end{figure}

In figure \ref{fig:lifshitz2} (right) we show the temperature dependence of $\chi(\mu)$ for suspended
graphene ($\lambda=1$). Here we chose a fixed time-discretization of $\delta_\tau=1/6~ \textrm{eV}^{-1}$, which
produces a nearly constant negative shift of $\chi$ (with only a very weak dependence on $\mu$) as the
leading discretization effect. Lattice sizes scale with $\beta$ such that the infinite volume
limit is obtained in each case. This requires larger systems at smaller temperatures, such that the
displayed curves correspond to $N=12,18,24$ respectively. Here we again observe a stronger peak at
$\mu\approx \kappa$, which grows as the temperature is lowered.
What is striking is that this increase receives sizable contributions from the disconnected part (shown in
figure \ref{fig:lifshitz3}, left), which is also unaffected by negative offsets from the
Euclidean time discretization.

Figure \ref{fig:lifshitz3} (right) shows the temperature dependence for the peak heights of
$\chi/\chi_\mathrm{con}/\chi_\mathrm{dis}$.
We have identified a range of $\beta =1/T$ between  $1.0$ eV$^{-1}$
and $3.0$ eV$^{-1}$ where a fit of the form
\begin{equation}
f_1(T)=a \ln\left(\frac{\kappa}{T}\right)+ b + c \, \frac{T}{\kappa}
\label{eq:logdivergence}
\end{equation}
to the full susceptibility is possible (it breaks down if one attempts
to include lower temperatures). More interestingly, however,
the same fit to $\chi_\mathrm{con}$ alone
is consistent with $a = 3/(\pi^2\kappa ) $ for $\beta \le 2.5$ eV$^{-1}$, and thus
basically fully agrees with the non-interacting tight-binding model.
At temperatures below $T\sim 0.15\, \kappa $ the contribution from
$\chi_\mathrm{con}$ suddenly drops however. This is contrasted by a rapid increase of the peak height of
 $\chi_\mathrm{dis}$. While $\chi_\mathrm{dis}$ is
negligible at high temperatures, it becomes the dominant contribution
at $T\sim 0.07 \, \kappa$. In fact, we find that for
$\beta \geq 2.5\textrm{ eV}^{-1}$ (corresponding to $T \le 0.15 \, \kappa$), $\chi_\mathrm{dis}^\mathrm{max} $
is well described by the model
\begin{equation}
f_2(T)=k \left|\frac{T-T_c}{T_c} \right|^{-\gamma}~,
\label{eq:powdivergence}
\end{equation}
with $\beta_c =6.1(5)\,\textrm{eV}^{-1}$ and $\gamma=0.52(6)$.
The emerging peak in $\chi^\mathrm{max}_\mathrm{dis}(T)$
around $\beta \approx 6$ eV$^{-1}$ is thus consistent
with a power-law divergence indicative of a thermodynamic phase
transition at non-zero $T_c$. All
attempts to model $\chi_\mathrm{dis}^\mathrm{max}(T)$ using a
logarithmic increase as in (\ref{eq:logdivergence}) were
unsuccessful, so that our conclusion seems qualitatively robust.

\section{Graphene with hydrogen adatoms}
\label{sec:vacancies}

Functionalization of graphene with hydrogen or other adatoms is a subject of interest, as it provides a way to create a tunable band gap
in graphene or to control its magnetic properties. The spatial
distribution of adatoms thereby plays a crucial role, with the question of stability (or instability) of adatom superlattices or other
configurations being of central importance. The smallness of the pairwise elastic interactions of hydrogen adatoms in graphene suggests that the
Ruderman-Kittel-Kasuya-Yosida (RKKY) contribution from conduction electrons dominates the inter-adatom interactions.
The influence of electron-electron interactions on these is quite strong in graphene, but is difficult to study
quantitatively with Density Functional Theory or similar methods.

In \cite{PhysRevB.96.165411} we carried out a HMC study of the RKKY interaction between hydrogen adatoms in graphene,
consistently taking into account inter-electron interactions. In the interacting tight-binding model the RKKY interaction is simply the fermionic
Casimir potential. For a pair of adatoms we calculate it as the free energy $\mathcal{F}_{xy}$ of the electrons on the graphene lattice with adatoms at
sites $x$ and $y$. In absence of inter-electron interactions we can simply compute the corresponding single-particle
energy levels $\epsilon_{xy}$ with adatoms and obtain $\mathcal{F}_{xy}$ up to an irrelevant constant $F_0$ from
\begin{eqnarray}
\label{free_energy_free}
 \mathcal{F}_{xy} = - T \sum_{\epsilon_{xy}} \ln\lr{1 + e^{-\epsilon_{xy} /T}} + F_0 ~.
\end{eqnarray}
With interactions, we instead calculate the differences $\Delta \mathcal{F} = \mathcal{F}_{x+l, y} - \mathcal{F}_{x, y}$ between free energies for adatom positions
which differ by a shift along one carbon-carbon lattice bond $l$, which we represent as
\begin{eqnarray}
\label{free_energy_integration}
\hspace{-10mm} \Delta \mathcal{F} = - T \int_0^1 d\alpha \, \partial_{\alpha} \, \log \mathcal{Z}_{\alpha} ,\quad
 \mathcal{Z}_{\alpha} = \int \mathcal{D}\phi_{x,\tau}\, e^{-S\lrs{\phi_{x,\tau}}}
|\det{M_{\alpha}\lrs{\phi_{x,\tau}}}|^2 .\label{interpolated_partition_func}
\end{eqnarray}
Here $M_{\alpha}$ linearly interpolates between fermionic operators with adatoms at positions $x$ and $y$ (at $\alpha = 0$) and $x+l$ and $y$ (at $\alpha = 1$). Differentiating the path integral (\ref{interpolated_partition_func}) for $\mathcal{Z}_{\alpha}$ by $\alpha$, we obtain
\begin{eqnarray}
\label{free_energy_integrand}
 \Delta \mathcal{F} = - 2 T \int_0^1 d\alpha \, \vev{\re\tr\lr{M_{\alpha}^{-1} \partial_{\alpha} M_{\alpha}}}.
\end{eqnarray}
The integral over $\alpha$ is calculated using the 6-point quadrature rule. The above is easily extended to more than two adatoms.

\begin{figure}
\begin{center}
\includegraphics[width=0.54\linewidth, trim=0 0 0 0]{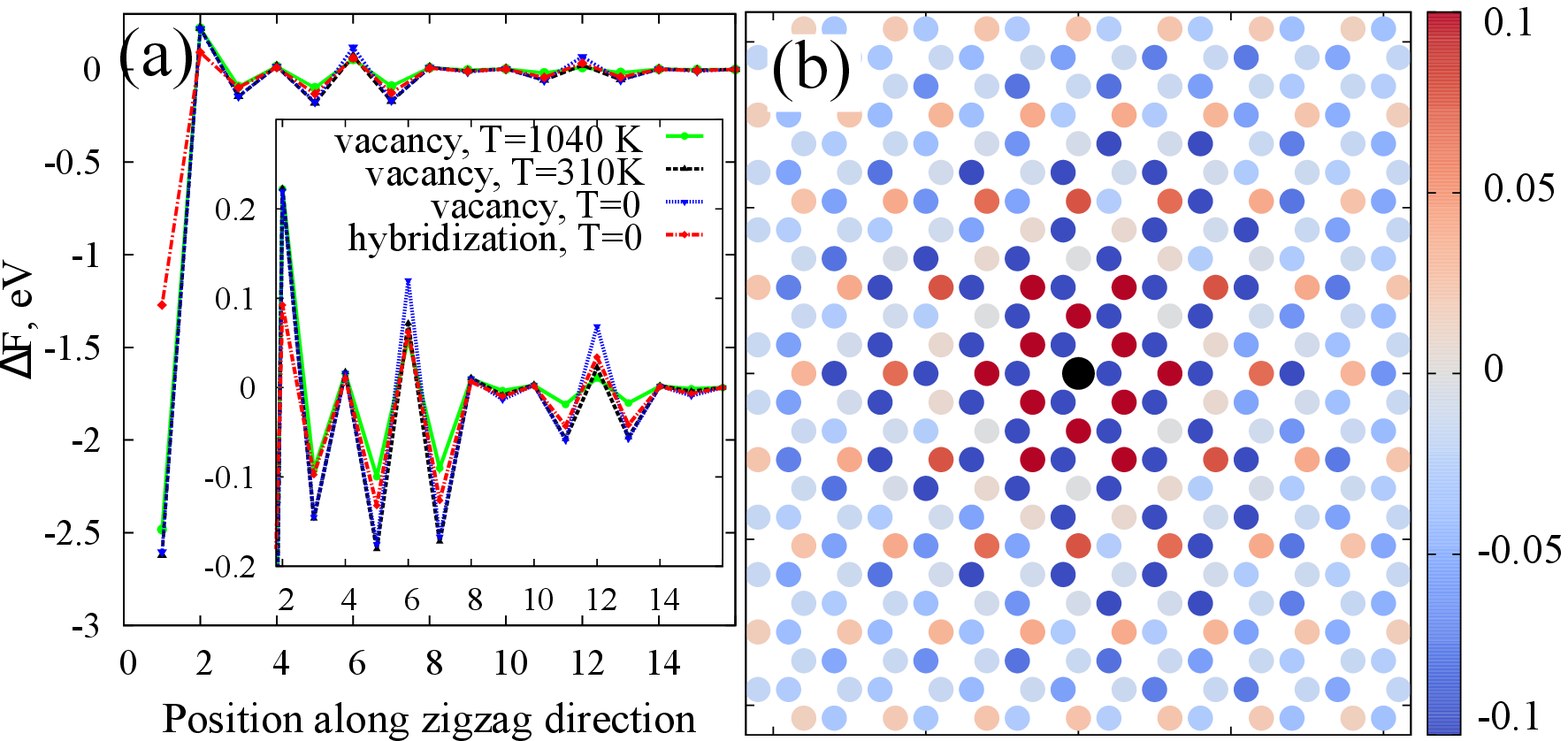}
\includegraphics[width=0.43\linewidth, trim=0 1.05cm 0 0]{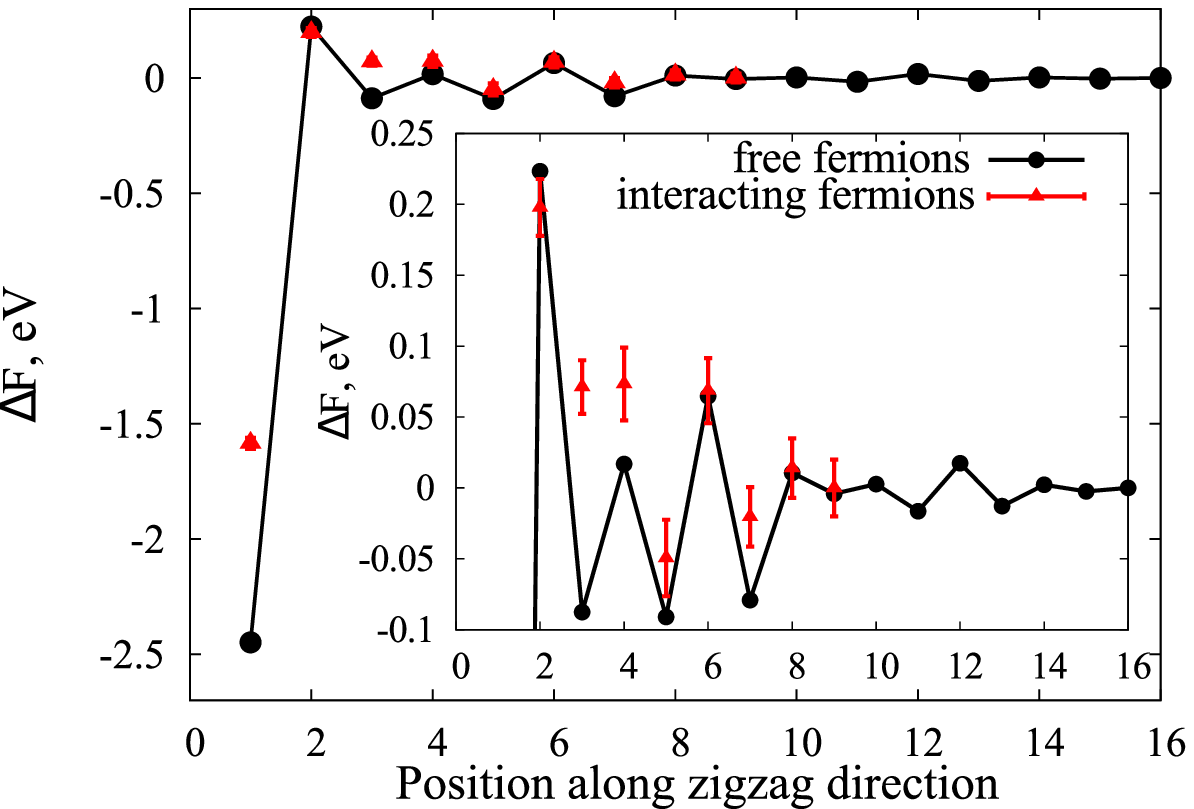}
\hspace{3mm}
\caption{Figures taken from \cite{PhysRevB.96.165411}. \emph{Left:} Interaction of two adatoms in free tight-binding model on $72\times72$ lattice. (a) profile along zigzag direction (zoomed version in the inset); (b) 2D profile of RKKY potential for non-interacting hybridization model at $T=310 \K$.
\emph{Right:}
Pairwise RKKY interaction in the interacting tight-binding model compared with non-interacting case. Zoomed version in the inset. Adatoms were modeled as vacancies.
}
\label{fig:vacancies1}
\end{center}
\end{figure}

We used two models of hydrogen adatoms: A simple vacancy model describing hydrogen adatoms
as missing lattice sites, and the full hybridization model, in which hybridization terms are added to the Hamiltonian.
As the hybridization model suffers from a sign problem, we use it only in the non-interacting limit.
It is used, among other things, to verify the validity of the vacancy model
and in cases where the effect of interactions can be modeled by SDW or CDW mass terms. In figure \ref{fig:vacancies1} (left)
we demonstrate that, without inter-electron interactions, the RKKY potentials are very similar for both models.
In all cases the pairwise RKKY interaction has well-known features: alternating signs for different sublattices and an order-of-magnitude enhancement at
some distances, at which the two adatoms induce midgap states with zero energy. We also demonstrate that relatively high temperatures
of $T=1040 \K$ do not affect these qualitative features.

In figure \ref{fig:vacancies1} (right) we illustrate the effect of inter-electron interactions on the RKKY potential along the zigzag direction
using the vacancy model. The potential is particularly strongly modified at distances of 3 and 4 C-C bonds, while at distances larger then 8-9 C-C bonds the
change of potential is too small to detect it with HMC. The main physical effect is that the local minimum at a distance of 3 bonds disappears and the potential
barrier between widely separated adatoms and the global minimum corresponding to a dimer configuration becomes harder to penetrate.

\begin{figure}
\begin{center}
\includegraphics[width=0.47\linewidth, trim=0 0 0 0]{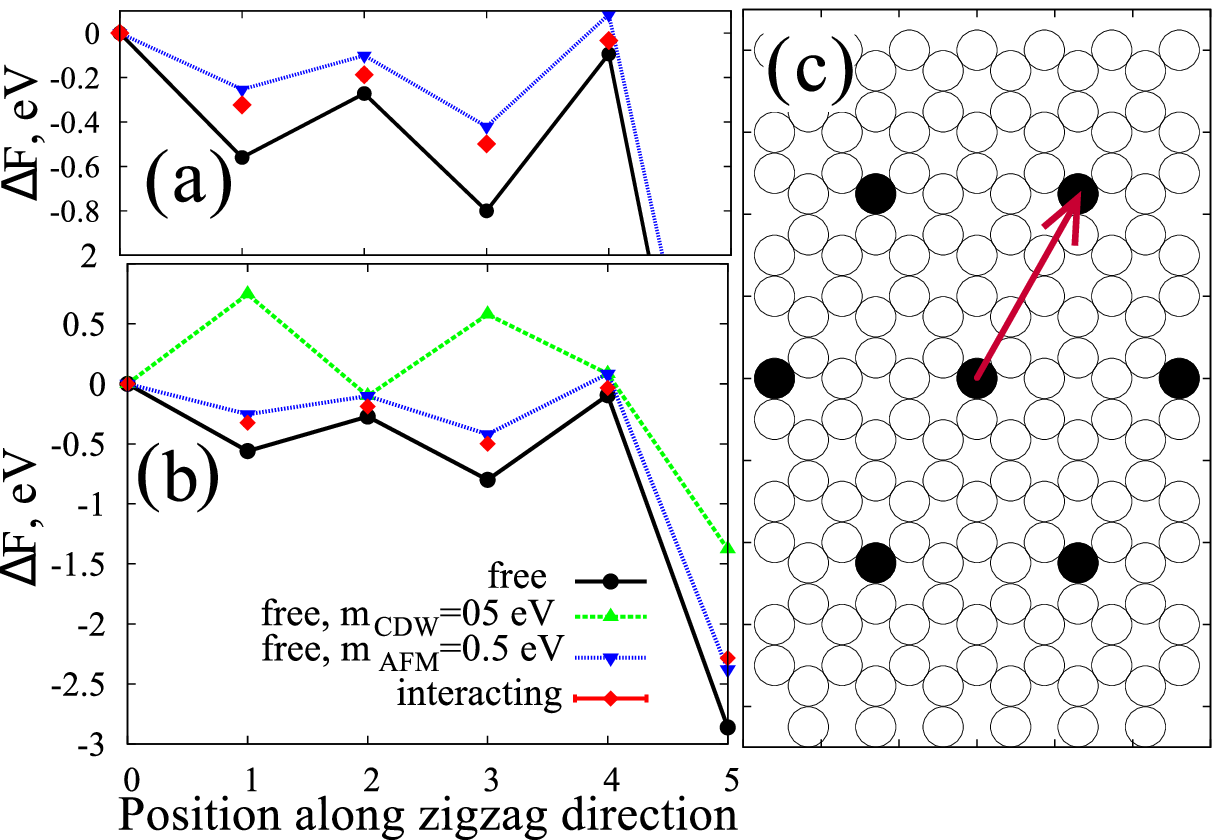}
\hspace{0.7cm}
\includegraphics[width=0.37\linewidth, trim=0 0 0 0]{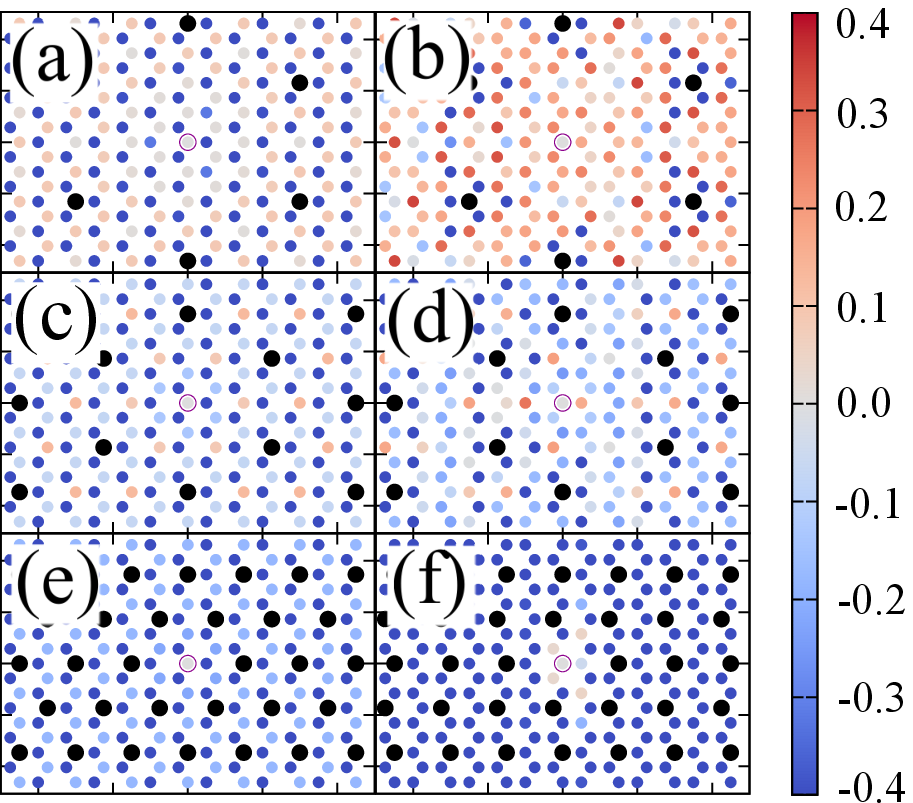}
\caption{Figures taken from \cite{PhysRevB.96.165411}. \emph{Left:} Free energy change of the superlattice system (vacancy model) upon displacement of a single adatom for $L=24$, $T = 0.09 \eV$ (zoom-in and overview); (c) superlattice structure with the zigzag profile used in figures (a) and (b) indicated by the red arrow.
\emph{Right:}
Change of free energy of superlattice systems upon the displacement of a single adatom. The fixed positions of other adatoms in the superlattices are
marked with black dots. All plots correspond to the non-interacting hybridization model with
SDW mass term at half-filling.}
\label{fig:vacancies2}
\end{center}
\end{figure}

We analyzed the stability of several regular adatom superlattices with respect to small
displacements of a single adatom. Figure \ref{fig:vacancies2} (left) shows one such system, with $5.56\%$
coverage of hydrogen adatoms populating only one sublattice, which is the case we studied in most detail.
Here the vacancy model is used in both the interacting and non-interacting case (to avoid a sign problem for the former and make a
direct comparison meaningful).
The overall
scale of the RKKY interaction is enhanced in comparison with pairwise interaction.
Inter-electron interactions do not change the RKKY potential qualitatively, despite inducing
a very large gap $\Delta \epsilon \sim 1 \textrm{eV}$ in the midgap energy band, and are modelled
quite accurately by adding a SDW mass term to the non-interacting tight-binding theory. A CDW mass
term on the other hand completely changes the RKKY potential and the locations of its minima.

We also addressed the dynamic stability of superlattice configurations with only one or both sublattices populated by
adatoms, by using the hybridization model without inter-electron interactions but with a SDW mass term.
Results are shown in figure \ref{fig:vacancies2} (right). We observe that the superlattices with adatoms on a single sublattice
at half filling are dynamically unstable in all cases considered, due to fact that a change of position of an adatom to the
opposite sublattice is energetically favourable. In contrast, superlattices of adatoms which equally populate both
sublattices are stable for low adatom concentration. Single-sublattice superlattices can be stabilized
by a chemical potential. This was also studied in \cite{PhysRevB.96.165411}. The main conclusion is that
a larger density of adatoms requires a larger chemical potential to stabilize it.

\section{Outlook}
\label{sec:outlook}

In conclusion, we note that there are of course several directions to continue each
of the projects presented here. The CPT scenario in graphene should certainly
be tested on larger system sizes, and the precise role of the short- and long-range
parts of the inter-electron potential investigated further. In particular, it would be instructive to repeat the analysis of 
section~\ref{sec:graphenephase} for the case of unscreened Coulomb interactions.

At the VHS one should study the competition between different ordered phases, in particular
of superconducting condensates, which can be expressed in a Nambu-Gorkov basis.
Genuine simulations at finite charge density are prevented by a sign problem, but
may be possible for small values of the chemical potential. In this context,
recently much progress has been made in applying the Lefschetz
thimble decomposition to the repulsive Hubbard model
\cite{Ulybyshev:2017hbs,Ulybyshev:2019fte,Ulybyshev:2019hfm}.
We are also in the process of adapting the
\emph{Linear Logarithmic Relaxation} method \cite{PhysRevLett.109.111601}, which generalizes
the Wang-Landau algorithm to systems with continuous degrees of freedom, to the repulsive Hubbard
model away from half filling \cite{PhysRevD.102.054502}.

We note that the chiral Gross-Neveu model is interesting in its own right, as an effective theory for chiral symmetry breaking in relativistic field theories, where it is better known as the Nambu-Jona-Lasinio (NJL) model. In $2+1$ spacetime dimensions, the attractive Hubbard model on the honeycomb lattice (which can be simulated at finite charge
density without a sign problem) defines a discretization of this theory similar to the Creutz-Borici action (which describes two degenerate flavors of chiral fermions, the minimum number allowed by the Nielsen-Ninomiya theorem), but which avoids
discretization errors produced by a vector-like anisotropy, typical for ``minimally doubled'' fermionic actions
on hypercubic lattices. This could be useful in the future e.g. to test the stability of inhomogeneous chiral
condensates beyond the mean-field level.

Finally, we point out that there are a great number of additional Hubbard-type models of interest
in condensed matter physics, which might possibly be targets for future HMC studies.
At present much attention is paid, for instance, to systems with flat bands in the spectrum, such twisted 
bilayer graphene \cite{Cao2018UnconventionalSI} or pseudospin-1 fermions on a dice lattice 
\cite{PhysRevLett.112.026402} (flat bands can also contribute to RKKY interactions \cite{PhysRevB.101.235162}).
Many such models might either be free of a fermion sign problem from the outset, or exhibit a sign
problem which turns out to be mild upon closer inspection, and hence tractable through various techniques. 
A detailed survey of such models is definitely of interest, but is beyond the scope of this work.

\section*{Acknowledgments}
This work was supported by the Deutsche Forschungsgemeinschaft  (DFG) under grants BU 2626/2-1 and SM 70/3-1. The work of P.B. was also supported by a Heisenberg Fellowship from the Deutsche Forschungsgemeinschaft (DFG), grant BU 2626/3-1. M.~U.~was also supported by the DFG grant AS120/14-1. D.~S.~and L.v.S.~ were also supported by the Helmholtz International Center for FAIR within the LOEWE initiative of the State of Hesse.

\section*{References}

\bibliographystyle{iopart-num}
\bibliography{Smith}

\end{document}